\RequirePackage{fix-cm}
\documentclass[smallextended]{svjour3}       
\smartqed  
\usepackage{graphicx}
\usepackage{listings}
\usepackage{tcolorbox}
\usepackage{xcolor}
\usepackage{colortbl}
\usepackage{amssymb} 
\usepackage{booktabs} 

\usepackage{multirow}   
\usepackage[utf8]{inputenc}  
\usepackage{makecell}

\usepackage{svg}

\usepackage{graphicx}
\graphicspath{ {./} }

\usepackage[hyphens]{xurl} 
\usepackage[colorlinks=true]{hyperref} 
\hypersetup{breaklinks=true}

\newcommand{\yes}{\checkmark}
\newcommand{\no}{\texttimes}

\begin{document} 
\title{Secure Coding with AI – From Detection to Repair}
\subtitle{Vulnerability Analysis and Fixing Using Large Language Models}

\author{Vladislav Belozerov  \and Peter J Barclay \and Ashkan Sami}

\institute{Vladislav Belozerov \at
              Edinburgh Napier University\\
              Edinburgh, Scotland \\
              \email{vladislav.belozerov@gmail.com}           
                      \and
           Peter J Barclay \at
              Edinburgh Napier University\\
              Edinburgh, Scotland\\
              \email{p.barclay@napier.ac.uk}  
           \and
           Ashkan Sami (Corresponding Author) \at
              Edinburgh Napier University\\
              Edinburgh, Scotland\\
              \email{a.sami@napier.ac.uk}  
}

\date{Received: date / Accepted: date}

\maketitle

\begin{abstract}

While several studies have examined the security of code generated by GPT and other Large Language Models (LLMs), most have relied on controlled experiments rather than real developer interactions. This paper investigates the security of GPT-generated code extracted from the DevGPT dataset and evaluates the ability of current LLMs to detect and repair vulnerabilities in this real-world context. We analysed 2,315 C, C++, and C\# code snippets using static scanners combined with manual inspection, identifying 56 vulnerabilities across 48 files. These files were then assessed using GPT-4.1, GPT-5, and Claude Opus 4.1 to determine whether these could identify the security issues and, where applicable, to specify the corresponding Common Weakness Enumeration (CWE) numbers and propose fixes. Manual review and re-scanning of the modified code showed that GPT-4.1, GPT-5, and Claude Opus 4.1 correctly detected 46, 44, and 45 vulnerabilities, and successfully repaired 42, 44, and 43 respectively. A comparison of experiments conducted in October 2024 and September 2025 indicates substantial progress, with overall detection and remediation rates improving from roughly 50\% to around 75–80\%. We also observe that LLM-generated code is about as likely to contain vulnerabilities as developer-written code, and that LLMs may confidently provide incorrect information, posing risks for less experienced developers.

\end{abstract}

\section{Introduction}
\label{intro}
The rapid advancement of technology and the increasing reliance on software in various industries have heightened the importance of secure software development. With the growing threat landscape, ensuring that software is robust and resilient against potential vulnerabilities is a critical concern for developers and organisations alike. Secure software coding practices have become an important aspect of the software development lifecycle, aiming to minimise security risks and protect sensitive data.

In recent years, the emergence of Generative AI as a software development tool, including use of GPT and other LLMs, has revolutionised the software development process. GPT, an LLM developed by OpenAI, has demonstrated remarkable capabilities in natural language processing, making it a valuable asset in various domains, including software coding. Although GPT can speed up writing custom software code, the security of such code is not clear. Here we explore the use of secure software coding practices when working with LLMs, examining the potential benefits and challenges associated with integrating AI tools into the software development process. 

Previous research has primarily focused on evaluating the capabilities of LLMs to generate functional code and their likelihood of introducing vulnerabilities. For instance, studies have assessed the performance of models like Codex and GitHub Copilot in code generation tasks, highlighting instances where insecure coding practices were evidenced \cite{Fu_Copilot_Weakness,Pearce_Asleep_at_the_Keyboard,Perry_Do_Users_Write_More_Insecure}. However, these investigations often utilised artificial prompts and scenarios crafted specifically for the experiments, which may not accurately reflect how developers interact with AI tools in practice. 

This paper addresses this gap by examining the security of code generated by GPT based on real developer interactions curated in the DevGPT dataset. This dataset provides a rich collection of developer-AI exchanges, capturing the nuances of how developers utilise GPT in practical coding tasks.  Our study is unique because it relies on real interactions between users and GPT, rather than on controlled lab experiments or specially crafted prompts. This hands-on approach gives us a more realistic view of GPT's behaviour, setting our work apart from other studies that rely mostly on lab-based or simulated code generation.

We selected for analysis code snippets in C, C++, and C\#, related languages which are widely used in various domains such as systems programming, game development, and enterprise applications. Code snippets were extracted from DevGPT \cite{DevGPT_Dataset}, a publicly available dataset, which contains developers interactions with GPT, including prompts, answers, and generated source code.

We evaluated the source code first with a set of static code scanners, then applied the LLMs to detect and fix the any issues identified. In October 2024 we used GPT‑4o to assess its ability to locate and fix vulnerabilities and, almost a year later in September 2025, we repeated the experiment with GPT‑4.1, GPT‑5, and Claude Opus 4.1.

To guide our investigation, we formulated the following research questions, shown below with a preview of our findings:

\begin{itemize}

\item \textbf{RQ1}: \emph{How secure is the code generated by GPT in the DevGPT dataset?}

\textbf{Answer}: Our analysis revealed that out of 2,315 code snippets, static scanners detected potential issues in 114 files. After manual review, we confirmed 56 vulnerabilities in 48 files, indicating that the code generated by GPT in the DevGPT dataset has security concerns.

\item \textbf{RQ2}: \emph{How effective are LLMs in detecting and correcting security issues when explicitly asked to do so?}

\textbf{Answer}: When explicitly prompted, GPT‑4o successfully detected 18 of the 32 confirmed vulnerabilities and resolved 17 of them, but it failed to recognise or fix the remaining issues. This demonstrates partial effectiveness while also highlighting the model’s limitations in consistently identifying and correcting security vulnerabilities.
The second experiment, conducted in September 2025, produced significantly better results: GPT‑4.1 detected 82\% of the problems and fixed 75\%, GPT‑5 detected and fixed 78.6\%, and Claude Opus 4.1 detected 80.4\% and fixed 76.8\%. All three models performed similarly, achieving roughly 80\%, a substantial improvement over the 53–56\% achieved by GPT‑4o in October 2024.

\item \textbf{RQ3}: \emph{Do developers provide more vulnerable code, or does GPT generate more insecure code during the interactions?}

\textbf{Answer}: 
In the first experiment only 10 vulnerabilities appeared in the developers' original prompts, while GPT‑4o introduced 22 issues, showing that the model produced more insecure code than the code supplied by humans. In the second experiment in September 2025, the distribution of flaws was roughly even: 25 of 56 issues were found in the user‑provided code, while 31 of 56 were in the GPT‑generated code.

\end{itemize}

The contributions of this paper are threefold:

\begin{enumerate}

\item \textbf{Empirical evaluation using real developer interactions}:  
We provide the first systematic assessment of LLM-generated code security using real-world developer prompts from the DevGPT dataset, addressing a gap in prior work that relied mainly on controlled experiments.

\item \textbf{Comparative analysis of modern LLM security capabilities}:  
We evaluated the effectiveness of GPT-4o, GPT-4.1, GPT-5, and Claude Opus~4.1 in detecting and repairing vulnerabilities, offering empirical evidence of model strengths and limitations in practical coding scenarios.

\item \textbf{Longitudinal assessment of LLM improvement}:  
By repeating the analysis at two points in time — October 2024 and September 2025 — we aimed to increase the likelihood that the findings would be generalisable regarding how code quality and LLM vulnerability-handling capabilities have evolved.

\end{enumerate}

Our findings underscore the importance of manual code review and caution against over-reliance on AI tools for secure code generation and analysis. We highlight the need for manual analysis and use of static checkers, noting that while AI tools can find vulnerabilities, they may also introduce them, and sometimes mislead developers regarding the security of the code under review.

The rest of the paper is organised as follows:  Section~\ref{sec:related_work} reviews related work on AI code generation and security. Section~\ref{sec:methodology} outlines the methodology. Section~\ref{sec:Data_exploring_and_collection} describes the data collection process, including data extraction and analysis procedures, while 
Section~\ref{sec:analysis_and_results} presents our results and discusses the implications. Section~\ref{sec:detection_and_correction} reviews LLMs' ability to detect and fix issues. Section~\ref{sec:threats_to_validity} presents threats to the validity of our work, and Section~\ref{sec:conclusion} concludes the paper with suggestions for future research.

\section{Related Work}
\label{sec:related_work}

Use of LLMs for coding, particularly GPT and Claude, has become 
a common practice, and a variety of recent papers have addressed 
the security implications of this development
\cite{Bakhshandeh_Using_Chat_GPT_as_Static_App_Security_tool,Cheshkov_Evaluation_of_GPT_Model_for_Vulnerability,Hamer_Just_another_copy_and_paste,Fu_How_Far_Are_We,Jamande_A_Pilot_Study_on_Secure_Code_Generation_with_GPT,Khoury_How_Secure_is_Code_Generated_by_GPT,Wu_Exploring_the_limits_of_GPT_in_software_sec}. Most published research focuses on either (1) 
assessing GPT-generated code for vulnerabilities, or (2) using GPT to detect security issues in existing code, approaches which we discuss 
in the following two subsections.

\subsection{Assessing GPT-generated Code for Vulnerabilities}

Bakhshandeh in \cite{Bakhshandeh_Using_Chat_GPT_as_Static_App_Security_tool} assesses GPT-3.5's performance in detecting vulnerabilities in Python source code, comparing it with three traditional static tools: Bandit \cite{bandit}, Semgrep \cite{semgrep}, and SonarQube \cite{sonarqube}. Here, 156 Python source code files were analysed using static analysers and GPT. Experiments demonstrated that GPT-3.5 outperformed the other tools in reducing false positives and negatives, making it a useful as a supporting tool. This study emphasised the importance of well-designed prompts to improve detection accuracy. 

Jamdade in \cite{Jamande_A_Pilot_Study_on_Secure_Code_Generation_with_GPT} explores GPT's ability to generate secure Node.js and PostgreSQL code for web applications, specifically addressing CWE-89 (SQL Injection) , CWE-79 (Cross-Site Scripting), CWE-93 (CRLF Injection), and CWE-200 (Exposure of Sensitive Information) vulnerabilities\footnote{CWE stands for Common Weakness Enumeration, a widely recognised classification system for software and hardware security weaknesses.}. Their paper concludes that GPT can be guided to generate secure code through specific prompts. The authors developed a university library system as a case study, and iteratively prompted GPT to improve the code; while improvements were observed, significant limitations in handling complicated scenarios still arose. These authors did not mention which GPT model version was used for experiments.

Khoury et al. in the study \cite{Khoury_How_Secure_is_Code_Generated_by_GPT} evaluate 21 simple programs generated by GPT across five languages (C, C++, Python, HTML, and Java) for specific vulnerabilities, including SQL injection, memory corruption, and cryptographic misuse. Results showed that GPT often generated insecure code, but could improve it when explicitly prompted about potential security problems. However, these ``fixed'' versions were still not robust against adversaries, often relying on naive mitigations like input validation or alphanumeric checks. The authors highlighted GPT's limited ability to generate secure code in advance, before the user explicitly asked it to generate secure code. 

Hamer, in~\cite{Hamer_Just_another_copy_and_paste}, compares Java code snippets generated by GPT with StackOverflow \cite{StackOverflow} answers using CodeQL for vulnerability analysis. The study analysed 108 snippets, revealing that GPT produced 20\% fewer vulnerabilities, but still propagated insecure patterns. GPT's snippets contained 248 vulnerabilities spanning 19 CWE types, compared to 302 vulnerabilities across 22 CWEs in StackOverflow. However, the 274 unique security issues across both sources highlight the risks of blindly relying on any of these platform for secure coding. Developers were advised to follow secure coding practices and carefully inspect code obtained from external sources.

Kharma et al. in \cite{Kharma_Security_and_Quality} performed a comprehensive study on modern LLMs -- including GPT-4, Claude-3.5, Codestral, Gemini-1.5, and Llama-3 -- by testing them on 200 coding tasks across Python, Java, C, and C++. Although many of the generated solutions were functional, the authors identified several hidden security risks, such as the use of outdated libraries or missing input validations. Their work highlights a common problem: while LLMs can speed up code production, developers must still run security checks to prevent potential exploits.

Manik in \cite{Motaleb_GPT_vs_DeepSeek} conducted a direct comparison between GPT-o1 and the newer LLM DeepSeek-R1 using an online platform Codeforces \cite{CodeForces} to verify 
the correctness of the code produced, and the static checker tools Pylint and Flake8 for quality analysis.  DeepSeek often solved the problems more accurately on its first attempt, but GPT's outputs tended to be shorter and more readable. The downside of GPT's more concise style was that it occasionally skipped security considerations, suggesting that features of convenience such as code brevity could overshadow vital checks.

Almanasra and Suwais \cite{Almanasra_Analysis_of_GPT_Generated_Codes} tested GPT-4o on 600 tasks that comprised 300 data-structure problems and 300 LeetCode \cite{leetcode_web} challenges, split across Python and Java implementations. Their results showed that while GPT-4 achieved a reasonable success rate and could handle even complex problems, it still left gaps in error handling and security measures. For example, the researchers found multiple instances of incomplete exception handling in both Python and Java solutions, emphasizing the requirement for human overview of the
resultant code.

The studies highlighted above show the growing importance of code generation and its potential for improving software security, with GPT showing promise but often generating insecure or incomplete outputs. 
Many researchers have highlighted the need for carefully
crafted prompts, and the importance of reviewing the code generated.

\subsection{Using LLMs as a Static Scanner}

Wu et al. in \cite{Wu_Exploring_the_limits_of_GPT_in_software_sec} compare GPT-3.5 and GPT-4.0 in ability to perform seven security tasks, including vulnerability detection and repair, debugging, symbolic execution, and fuzzing. Using benchmark datasets and manual test cases, the study shows GPT-4 significantly outperforms GPT-3.5 in accuracy and effectiveness. 
GPT demonstrated strong capabilities in structured tasks like vulnerability detection or assembly code decompilation, but has difficulties with longer code contexts, binary code, and real-world complexities. 

Cheshkov in \cite{Cheshkov_Evaluation_of_GPT_Model_for_Vulnerability} 
evaluates GPT-3 and GPT-3.5-turbo for vulnerability detection in Java code. Using a dataset of vulnerable and patched files, these models were assessed on binary and multi-label classification tasks for specific CWE types. The findings revealed significant limitations of GPT, poor detection performance and biases favouring patched code. The study concludes that GPT-based models are currently inadequate for effective vulnerability detection without further refinement.

Fu et al. in \cite{Fu_How_Far_Are_We} evaluate GPT’s performance on four software vulnerability related tasks: prediction, classification, severity estimation, and repair using real-life datasets of vulnerability-fixing commits from open-source C/C++ projects. GPT underperforms compared to specialised models such as CodeBERT \cite{CodeBert}, highlighting the need for domain-specific fine-tuning. Despite its scale, GPT struggles without adaptation for vulnerability contexts and demonstrates limited effectiveness in detecting vulnerabilities.

Kholoosi, Babar, and Croft in \cite{Kholoosi_A_Qualitative_Study} pursued a different angle by looking at real-world impressions from security professionals who tried GPT (using version 3.5) for tasks such as vulnerability detection and penetration testing. After collecting information from X/Twitter posts, the authors ran their own controlled experiments to see if GPT's advice was trustworthy. The results were mixed: participants found GPT's suggestions helpful for brainstorming or retrieving general security knowledge, but the authors' tests showed it frequently overlooked subtle but important edge cases. This raises questions about whether GPT can serve as a fully reliable “scanner” without additional tools or specialised training data.

Sajadi et al. in \cite{Sajadi_Do_LLMs_Consider_Security} went further by testing Claude 3, GPT-4, and Llama 3 on real Stack Overflow snippets containing known vulnerabilities in Python or JavaScript. According to the authors, none of the models consistently issued security warnings in response to these faulty code examples unless the prompt explicitly asked about security. However, when the models did detect a problem, they produced surprisingly detailed explanations or fixes. These findings confirm that LLMs could grow into more robust scanning tools if provided with context-rich instructions with a strong focus on security from the outset.

Yu et al. in \cite{X_Fight_Fire_With_Fire} evaluated GPT's self-verification ability using synthetic benchmark datasets, relying on controlled experiments with GPT-3.5 and GPT-4. 
In their work, they found that GPT frequently misjudges its own code, often labelling incorrect or vulnerable programs as correct or secure. They also observed self-contradictory behaviour, where the model later disagrees with its earlier assessment, demonstrating the significant limitations of GPT’s self-verification ability.

In contrast to earlier work, our study analyses real-world developer interactions from the DevGPT dataset, which includes C, C++, and C\# code originally generated by GPT-3.5 and GPT-4 during real user interaction sessions. We apply multiple static scanners, namely Flawfinder, Snyk, Semgrep, and Cppcheck and assess the vulnerability detection and fixing abilities of GPT-4o, 4.1, 5, and Claude Opus 4.1. This approach provides a complementary perspective to benchmark-based work and allows us to investigate some new aspects of the problem in realistic coding scenarios.

\subsection{Summary of Prior Research}

All these studies show mixed results, with GPT able to detect a range of vulnerabilities, but proving inadequate to give confidence that all problems would be identified.

The overview in Table \ref{tab:papers_info} includes data for all the  papers reviewed as well as this the current paper. The shortened column names in this table have the following meanings: ``Lang'' means ``Programming Language'', ``Detect'' means ``Is security issues detection ability tested?'', and ``Generat'' means ``Is code generation ability tested?''.
The data presented in this table highlight how our paper effectively addresses existing gaps in the previous research. While many previous studies relied predominantly on specifically generated or manually reviewed datasets, our paper distinguishes itself by using the DevGPT dataset, which provides real-life code examples rather than purely synthetic or manually curated sources. Moreover, it uniquely uses a comprehensive set of static scanners including Semgrep, Flawfinder, Snyk, and CPPCheck, and investigates their detection capabilities across multiple related languages, namely C, C++, and C\#. 
Additionally, we compare the performance of several LLMs in detecting security issues against the results from these static 
scanners.

\begin{table}[htbp]
\caption{Literature review paper comparison}
\begin{center}
\begin{tabular}{p{1.0cm} p{1.9cm} p{2.5cm} p{2.5cm} p{1.0cm} p{1.0cm}} 
\hline
\textbf{Paper} & \textbf{Lang.} & \textbf{Static scanners} 
& \textbf{Real-life code example or lab generated} & \textbf{Detect.} 
& \textbf{Generat.}\\

\hline
\textbf{This paper}	& C/C++, C\# & Semgrep, Flawfinder, Snyk, CPPCheck	& DevGPT dataset &	Yes	& Yes
\\

\hline
Bakhshan deh \cite{Bakhshandeh_Using_Chat_GPT_as_Static_App_Security_tool}
 &	Python	& Bandit, Semgrep, SonarQube	& Real datasets: securityEval, PyT	& Yes	& No
\\

\hline
Jamdade  \cite{Jamande_A_Pilot_Study_on_Secure_Code_Generation_with_GPT}
	& TypeScript &	No &	Specifically generated	& Yes	& Yes

\\

\hline
Khoury \cite{Khoury_How_Secure_is_Code_Generated_by_GPT}
	& C, C++, Python, HTML, and Java	& No, manual review & 	Specifically generated	& Yes	& Yes
\\

\hline
Hamer \cite{Hamer_Just_another_copy_and_paste}
	& Java & 	CodeQL	& Specifically generated, StackOverflow used to collect examples of code produced by humans	& No	& Yes
\\

\hline
Kharma \cite{Kharma_Security_and_Quality} 
 &C, C++, Java, and Python &	SonarQube	& Specifically generated	& No &	Yes
\\

\hline
Manik \cite{Motaleb_GPT_vs_DeepSeek}
 & Python &	Pylink, Flake8	& Specifically generated &	No	& Yes
\\

\hline
Almanasra \cite{Almanasra_Analysis_of_GPT_Generated_Codes} 
 & Java, Python	& Leetcode to test memory and speed efficiency	& Specifically generated &	No	& Yes
\\

\hline
Wu \cite{Wu_Exploring_the_limits_of_GPT_in_software_sec}
 & C, C++, Java, Go, and Python	& Manual checks	& Specifically generated &	Yes	& Yes
\\

\hline
Cheshkov \cite{Cheshkov_Evaluation_of_GPT_Model_for_Vulnerability} &  Java	& Manual checks	& Samples from GitHub &	Yes	&  No
\\

\hline
Fu \cite{Fu_How_Far_Are_We}
 & C/C++	& AIBugHunter, CodeBERT, GraphCodeBERT, and VulExplainer were used as static scanners	& Real life datasets Big-Vul and CVEFixes &	Yes	& Yes
\\

\hline
Kholoosi \cite{Kholoosi_A_Qualitative_Study} 
 & 12 different languages	& No &	National Vulnerability Database (NVD) &	Yes & No
\\ \hline
Sajadi \cite{Sajadi_Do_LLMs_Consider_Security}
 	& Python and JavaScript	& CodeQL &	Stack Overflow data &	Yes	&  No
\\ \hline

Yu \cite{X_Fight_Fire_With_Fire} & 13 programming languages including Python, C++, Java, C\#, Go, JS, Kotlin, etc. & CodeQL &  Specifically generated & Yes & Yes  \\ \hline       

\end{tabular}
\label{tab:papers_info}
\end{center}
\end{table}

Overall, previous studies leave gaps in addressing multilingual contexts and in real‑world validation. Moreover, they focus only on the code itself, neglecting the interaction between developers and LLMs that produces and analyses that code. In contrast, our work analyses code from the DevGPT dataset using a structured data pipeline that combines static analysis, manual validation, and analyses LLM responses to user requests aimed at detecting and fixing security issues.

Our study provides a systematic approach using static analysis with multiple scanners, manual validation, assessing the
LLMs' repair capabilities for C, C++, and C\#, providing practical insights into selected LLMs' effectiveness in fixing security problems. An additional insight from our study is that the experiments were conducted approximately one year apart, allowing us to observe a significant improvement in the LLMs' performance over time.

\section{Methodology}
\label{sec:methodology}

\begin{figure}[h]
    \centering
    \caption{Dataflow diagram}
    \includegraphics[width=10.96cm, height=7.68cm]{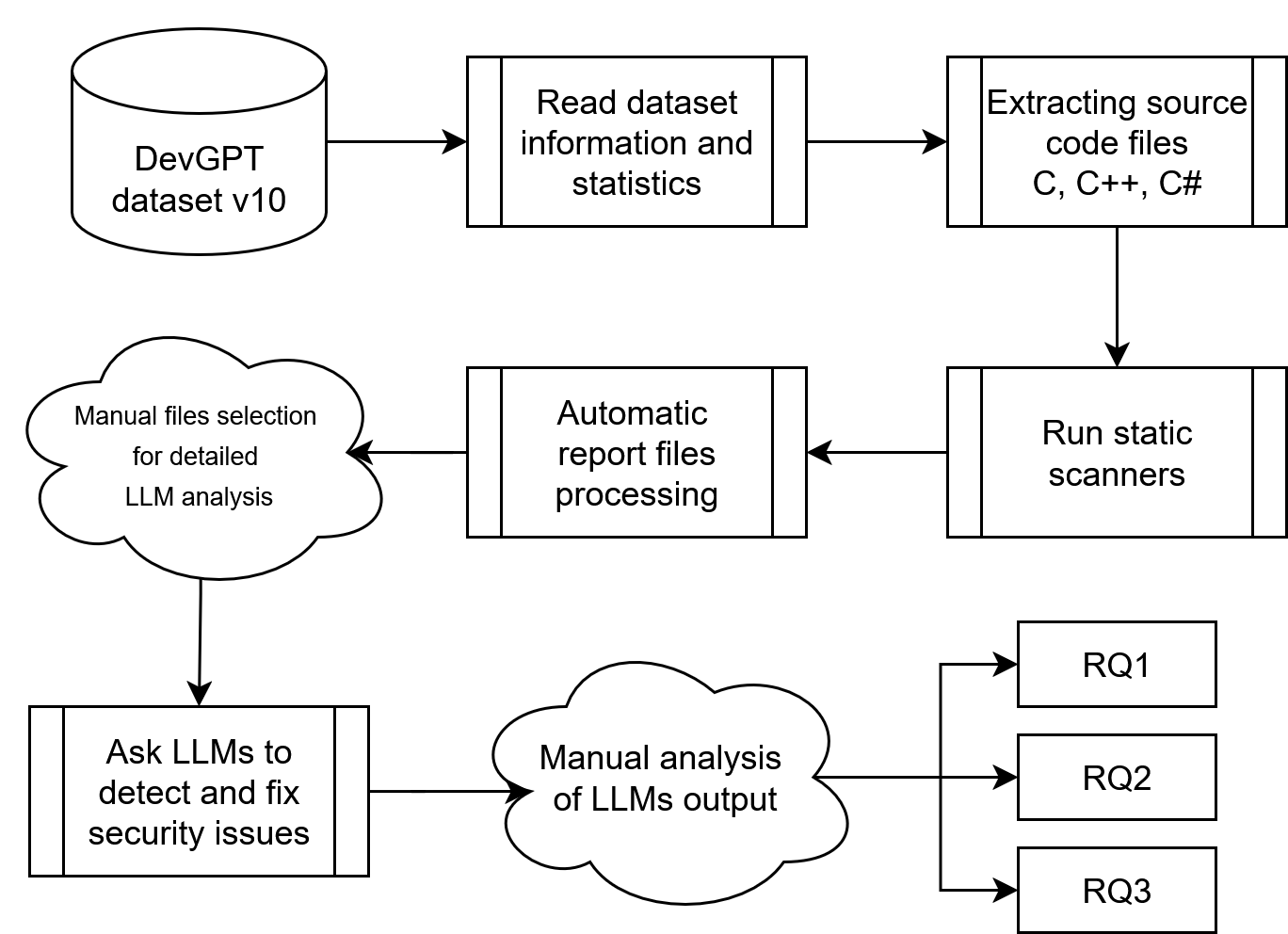}
    \label{fig:dataflow_diagramm}
\end{figure}

The overall workflow of the study including dataset preparation, static analysis, automatic processing, LLM evaluation, and manual verification is presented in Figure \ref{fig:dataflow_diagramm}.

We begin by processing the DevGPT dataset, using Python scripts to extract high-level information such as dataset composition, basic statistics, and the distribution of programming languages. We then identify and extract all ``Sharing'' folders containing C, C++, and C\# source files for further analysis. These files are analysed with a set of static scanners, namely Flawfinder, Semgrep, Snyk, and Cppcheck, and the resulting reports are automatically collected, parsed, and transformed using custom Python scripts. After a brief review of the aggregated reports, a dedicated selection algorithm identifies the most relevant source files for in-depth LLM analysis. For these selected files, we queried multiple LLMs (GPT-4o in 2024 and GPT-4.1, GPT-5, and Claude Opus 4.1 in 2025) to detect and repair potential security issues. Finally, we manually examine each model’s output, verifying whether the original code contained an actual vulnerability, whether any static-scanner findings were false positives, and whether the LLM correctly detected and fixed the issue. This process allows us to answer the three research questions informing this study.

The proposed methodology was implemented through a sequence of data preparation, static analysis, LLM-based code evaluation, and manual verification. We began with dataset inspection and preprocessing, where the source repository is explored, code snippets are extracted, and the dataset is prepared for the analysis. This included transforming the raw data into a structured format of code snippets suitable for both static scanners and LLM-based assessment experiments.

We then proceeded with automated static analysis, where multiple scanners were applied to detect potential vulnerabilities. The outputs from these tools were collected and processed to interpret the results. In this stage, based on the detection capabilities of traditional static analysis tools, we identified potential vulnerabilities in the code snippets.

Following this, we manually reviewed and summarised the scanner outputs to ensure relevance and suitability for deeper analysis. Next, we investigated
the effectiveness of the LLMs, using GPT-4o on source code
excerpts from the curated dataset of 2024, 
and subsequently GPT-4.1, GPT-5, and Claude Opus 4.1 for the 2025 dataset, to assess their capabilities to detect and repair security vulnerabilities in the selected code samples. The analyses were performed on both user-provided code in the prompts and LLM-generated code snippets.  
Lastly, every model output was manually assessed to confirm whether detected vulnerabilities reported by static scanners are presented also by the LLMs, and whether the LLM-proposed fixes were valid and complete. This comprehensive evaluation allows us to compare static analysis tools and LLM-based approaches and to address the research questions.

\section{Data Preprocessing}
\label{sec:Data_exploring_and_collection}

The data for our analysis were taken from the publicly available DevGPT \cite{DevGPT_Dataset} dataset, which contains information on developer interactions with GPT, including prompts, answers, and generated source code. This information was collected by parsing GPT discussions which were shared on Hacker News \cite{Hacker_news}, GitHub \cite{Github_web} issues, GitHub pull requests, GitHub commit messages, and GitHub source code files. GitHub and HackerNews both provide APIs which were used to retrieve the data for DevGPT.

DevGPT was selected because it is an open access dataset with a permissive license and contains GPT conversations from a variety of sources. It contains the actual output of AI tools, making it a good source for analysing how well these tools perform in a real-world programming tasks. In addition, DevGPT includes code snippets from various domains such as web development, tool development, data processing, and algorithm design.

The DevGPT dataset is organised into six sections, each containing six JSON files covering GitHub Issues, Pull Requests, Discussions, Commits, Code Files, and Hacker News  threads. 

DevGPT version 10 was used as the dataset, specifically the snapshots \texttt{snapshot\_20231012} and \texttt{snapshot\_20240514}.
These snapshots were created in October 2023 and May 2024 respectively.
This means there is at least a four-month gap between the date when the source code was collected in DevGPT and when it was inspected by the LLMs. All code samples in DevGPT were originally generated by GPT-3.5 and GPT-4.

Python was used to develop an auxiliary toolset for file extraction and data analysis. GitHub repository containing the developed software tools \cite{Tools_GitHub}. The DevGPT dataset was analysed using these tools, and the dataset contained the items shown in Table \ref{tab:dataset_info}.  The most popular languages in the DevGPT dataset are presented in Table \ref{tab:prog_lang}.

\begin{table}[h]
\caption{DevGPT dataset information with breakdown by dataset snapshots}
\begin{center}
\begin{tabular}{  p{3cm} p{2cm} p{2cm} p{2cm} } \hline

\textbf{Metric name} & \textbf{DevGPT snapshot \_20231012} & \textbf{DevGPT snapshot \_20240514} & \textbf{Total number of items} \\ \hline

Separate GPT interaction cases (sharings) & 4733 & 5645 & 10378 \\ \hline
Conversation steps & 27093 & 23835 & 50928 \\ \hline
Source code files & 19106 & 19907 & 39013 \\ \hline

\end{tabular}
\label{tab:dataset_info}
\end{center}
\end{table}

\begin{table}[h]
\caption{Programming languages in DevGPT dataset with breakdown by dataset snapshots}
\label{tab:prog_lang}
\begin{center}
\begin{tabular}{  p{3cm} p{2cm} p{2cm} p{2cm} }
\hline

\textbf{Programming language} & \textbf{DevGPT snapshot \_20231012} & \textbf{DevGPT snapshot \_20240514} & \textbf{Total number of files} \\ \hline
 
Python &	3041 & 3864 & 6905 \\ \hline
Java script	& 2401 & 2476 & 4877 \\ \hline
bash	 & 2166	& 1750 & 2166\\ \hline
HTML & 	1055	 &1050 & 2105\\ \hline
Unix shell	& 751 & 	1203 & 1954 \\ \hline
Java	& 461 &	833 & 1294 \\ \hline
C++	& 544 &	700 & 1244\\ \hline

C\#	 &  898	& 198 & 1096 \\ \hline
CSS	& 641 & 297 & 938  \\ \hline
Swift	&  487 & 72 &559\\ \hline
Rust & 232 & 317 & 549 \\ \hline
Go	& 354 & 169 & 523\\ \hline
C	& 144 &	210  & 354    \\ \hline

\end{tabular}

\end{center}
\end{table}

This study focuses on C\#, C, and C++  as these languages are widely used in various domains such as systems programming, game development, enterprise applications, critical infrastructure control and security applications. By focusing on the types of security vulnerabilities typical for C-family languages, this research aims to provide insight into issues that may be expected to commonly appear in AI-generated code in these application areas.

\begin{table}[h!]
\caption{Number of code lines for selected programming languages with breakdown by dataset snapshots}
\label{tab:code_lines}

\centering
\begin{tabular}{l c c c c c c }
\hline

\multirow{3}{*}{\textbf{Language}} & \multicolumn{2}{c}{\textbf{DevGPT}} & \multicolumn{2}{c}{\textbf{DevGPT}} & \multicolumn{2}{c}{\textbf{Total}} \\ 

                          & \multicolumn{2}{c}{\textbf{snapshot\_20231012}} & \multicolumn{2}{c}{\textbf{snapshot\_20240514}} &  \\ \cline{2-7}
                          
                          & \textbf{Files} & \textbf{Lines} & \textbf{Files} & \textbf{Lines} & \textbf{Files} & \textbf{Lines} \\ \hline
C                         &  121    &  1784    &  204    &  5665    &  325    &  7449    \\ \hline
C++                       &   427 &	7559 &	563 & 12310   &   990   &   19869   \\ \hline
C\#                       &   829 &	14824 &	171 & 2801   &   1000   &  17625    \\ \hline
\end{tabular}
\end{table}

The number of files and number of lines of source code used for analysis of each programming language can be found in  Table \ref{tab:code_lines}. We analysed a total of 2,315 source files, which included 1000 C\# files, 990 C++ files, and 325 C files. C\# was the most prevalent language, accounting for 1000 files and 17,625 lines of code, followed by C++ and C with decreasing numbers. Source code file statistics were generated using the \texttt{cloc} utility~\cite{cloc_tool}.

Another Python tool was developed to iterate over the dataset and extract data in the form of files suitable for further analysis. The code samples were organised into a folder tree with the following structure:\\

\texttt{Code -> JsonFile\_Folder -> Source\_<Name> -> Sharing\_<Name\_Date> -> Conversation\_<N>} \\

Where:

\begin{itemize}
\item \texttt{<Name>} -- is specific name of the source.
\item \texttt{<Name\_Date>} -- is the name and date of the conversation in each sharing.
\item \texttt{<N>} -- is the number of the conversation.
\end{itemize}

Each ``\texttt{Sharing\_<Name\_Date>}'' folder contains information in the file called \texttt{sharing\_info.txt} with the title, date of interaction, the URL to the GPT conversation, the GPT model name, and number of conversations. The term ``sharing'' is used here as defined by DevGPT documentation, referring to a shared session of interaction between the user and GPT. In other words, it represents a conversation between the user and GPT on a selected topic.

The entire conversation in HTML format is stored in a file with the name \texttt{html\_content.html} within the \texttt{Sharing\_<Name>} folder. Extracting data into files was necessary for two reasons: first, to store the source code as files, thereby enabling the use of static scanners; secondly, for convenience, as navigating files on disk and exploring the dataset is easier than working with JSON content.

\section{Data Processing and Static Analysis Results}
\label{sec:analysis_and_results}
\subsection{Static Scanners Analysis}

As discussed in \cite{Goseva_On_the_capability_of_static_code_analysis} 
and \cite{Kulenovic_Survey_static_code_analysis}, static code scanners are essential in software development for identifying security vulnerabilities early in the development lifecycle, thereby reducing the cost and effort required to fix them. Scanners can automate the detection of common security issues, ensuring consistent application of security standards across the codebase. Numerous scanners are available on the market. Four well-known scanners were selected, based on the criteria of offering a free or trial version and supporting the chosen programming languages. The selected scanners presented in Table \ref{tab:scanners}.

\begin{table}[h]
\caption{Static scanners selected for analysis}
\label{tab:scanners}
\begin{center}

\begin{tabular}{ >{\raggedright\arraybackslash}p{1.8cm} 
                 >{\raggedright\arraybackslash}p{1.8cm} 
                 >{\raggedright\arraybackslash}p{2.5cm} 
                 >{\raggedright\arraybackslash}p{1.8cm} 
                 >{\raggedright\arraybackslash}p{1.8cm} }
\hline

\textbf{Scanner} & \textbf{Version} & \textbf{Programming Languages}& \textbf{Is free or trial available}& \textbf{Web site}\\
\hline
Snyk &	1.1294.0 &	C\#, C, C++	& Yes &	\cite{snyk} \\ \hline
Semgrep	& 1.87.0 &	C\#, C, C++ &	Yes	 & \cite{semgrep} \\ \hline
FlawFinder & 	2.0.19 & 	C, C++ & 	Yes & 	\cite{flawfinder}  \\ \hline
CppCheck & 	2.14.0 & 	C, C++ & 	Yes & 	\cite{cppcheck} \\ \hline
\end{tabular}
\end{center}
\end{table}

\begin{table}[h!]
\caption{Static scanning results}
\label{tab:scan_results}
\centering
\begin{tabular}{c  c  c  c c}
\toprule
\textbf{Scanner} & \textbf{Number of detects} & \multicolumn{3}{c}{\textbf{Detect severity}} \\ 
\cmidrule(lr){3-5}
& & \textbf{Errors} & \textbf{Warnings} & \textbf{Notes/Info} \\ \midrule
Snyk & 153 & 1 & 150 & 2 \\ 
Semgrep & 56 & 0 & 56 & 0 \\ 
Flawfinder & 679 & 20 & 23 & 633 \\ 
CppCheck & 71 & 0 & 71 & 0 \\ 
\hline 

Total & 959 & 21 & 300 & 635 \\
\bottomrule
\end{tabular}
\end{table}

All the files extracted from the DevGPT dataset were scanned using these four scanners, and a scanning report was generated for each scanning run.   The code snippets were extracted from the dataset as separate files before scanning with these static analysis tools. It should be noted that only the raw source code, without any conversational text, was submitted to prevent inaccurate scanner results. The source code was not compiled or modified before scanning, and many fragments could not be compiled due to syntax errors or incomplete content.

Raw scanning results are presented in Table~\ref{tab:scan_results}, where we observe that Flawfinder found 679 possible issues, almost 2.5 times more than the all other scanners combined. However, most of these detections had minor ``Notes/Inf'' severity, and only 43 more severe issues were detected. For all static scanners except CppCheck, reports were generated in \texttt{SARIF} format, a popular JSON-based format developed by OASIS \cite{Sarif_format}. Since CppCheck does not support \texttt{SARIF} generation, its report was generated as a plain text file.
Tables~\ref{tab:scan_results} and \ref{tab:defects_per_line} contain information from DevGPT snapshots \texttt{snapshot\_20231012} and  \texttt{snapshot\_20240514}.  The former corresponds to October 2023, and the latter corresponds to May 2024.  Statistics on defects per line of code presented in Table~\ref{tab:defects_per_line}. 

\begin{table}[h]
\caption{Number of detects statistics per line of code}
\label{tab:defects_per_line}
\begin{center}
\begin{tabular}{
>{\raggedright\arraybackslash}p{1.2cm} 
>{\raggedright\arraybackslash}p{2.0cm} 
>{\raggedright\arraybackslash}p{0.8cm} 
>{\raggedright\arraybackslash}p{1.3cm} 
>{\raggedright\arraybackslash}p{1.0cm} 
>{\raggedright\arraybackslash}p{1.5cm} 
>{\raggedright\arraybackslash}p{1.5cm} }

\hline
\textbf{Scanner} & \textbf{Languages}& \textbf{Files}& \textbf{Lines of code}& \textbf{Detects} & \textbf{Number of files per detect} & \textbf{Lines of code per detect}\\
\hline
Snyk &	C\#, C, C++	& 2315 & 44943 & 153 & 15.1 & 293.7 \\ \hline
Semgrep	& C\#, C, C++ &	2315 & 44943 & 56 & 41.3	& 802.6 \\ \hline
FlawFinder & C, C++ & 	1315 & 	27318 & 679	& 1.9 &	40.2 \\ \hline
CppCheck & 	C, C++ & 	1315 & 	27318 & 71 & 18.5 &	384.8 \\ \hline

\end{tabular}
\end{center}
\end{table}
\subsection{Review of Initial Detections and Report File Transformation}

\label{subsec:review_and_files_selection}

Since the \texttt{SARIF} format contains a level of detail unnecessary for our analysis, and has a  complicated internal structure, we decided to create a simplified JSON-based format to store information
from the static scanner reports. The goal of this format is to retain the essential information from the CppCheck report in a structured form, at the same time avoiding the complexity of the \texttt{SARIF} format. 

To ensure all reports were in a unified format, the other scanner reports were also converted to the same customised JSON format. At this stage, we had four JSON report files in the same format from four different scanners. Having the reports in a consistent format simplifies analysis and allows us to use the same approach for all selected static scanners. Each detection in a scanner report represents a potential security issue; using all the report data, we initiated a review and selection process, as some detections might be false positives or perhaps very minor issues.

With the help of the another custom Python script, we analysed all the JSON reports from each scanner and outputted the issues, grouped by ``Sharing'' folder name. As noted, a ``Sharing'' folder represents a single discussion between GPT and a user, containing multiple steps with prompts and responses.  

Table \ref{tab:scan_results} presents the number of files and the number of  detects for each scanner. After this stage, we performed a manual selection yielding a reduced set of files for more detailed analysis; the selection process is described below. 

Based on results from the static scanner analysis stage, the most interesting files were then subjected to
separate manual review by software experts.
The file selection process proceeded as follows:

\begin{itemize}
\item Initially, after running the static scanner analysis, we obtained a list of files where scanners had detected issues.
\item To be included in the analysis, each file must have at least one issue detected by at least one static scanner. This is a mandatory requirement for inclusion.
\item A file is selected if multiple scanners detected an issue on the same line. A special Python script was used to match detection information.
\item Using this Python tool, we grouped files from the static scanner reports by ``Sharing''.
\item  Only one file was selected from each ``Sharing''. A DevGPT ``Sharing'' can have multiple conversation steps, and we usually take only one file, typically from the last step of discussion.
\item The resulting list of files was then analysed using software tools to detect duplicated paths and remove duplications.

\item  Two snapshots from DevGPT were included to the analysis
(these were \texttt{snapshot\_20231012} and \texttt{snapshot\_20240514)}.
Files which were already covered in the first snapshot were excluded from the second set -- each ``Sharing'' has an associated URL field, which allows reliable automatic exclusion based on this parameter, using a Python tool. This step is necessary because DevGPT snapshot updates are not incremental, and some \texttt{snapshot\_20231012 files} are missing from \texttt{snapshot\_20240514}.

\item Finally, we obtained two unique file lists for analysis corresponding to the two selected DevGPT snapshots.
\item Only files from DevGPT \texttt{snapshot\_20231012} were used in the earlier \mbox{October} 2024 experiment. 

\end{itemize}

After completing this process, in October 2024 we identified 64 relevant files from DevGPT \texttt{snapshot\_20231012}. For the September 2025 analysis, we investigated 114 files selected from DevGPT snapshots 
\texttt{snapshot\_20231012} and \texttt{snapshot\_20240514}. 

\subsection{Manual Analysis of Scanner Output} 

The results for all 114 files were manually reviewed by the first author and the findings were discussed and validated by the second and third authors. After the selection process, we confirmed 56 security issues in 48 files. Other detections were marked as false positives. For example, in several cases scanners were triggered by a simple \texttt{printf()} function call with a \texttt{const char*} argument. Another case comprised a \texttt{memcpy()} call where the size check was done on a different line, before the \texttt{memcpy()} call, and some scanners were triggered by reading and returning \texttt{Boolean} values from functions without validation, which is usually unnecessary for such values.

The CWE distribution of the results is presented in Table \ref{tab:cwe_by_type}. A total of 66 different types of CWEs were identified, which exceeds the 56 vulnerabilities because, in some cases, the same vulnerability results in two CWEs being assigned to the same line of code. In the third column, we can see whether the issue is included in the 2024 CWE ``Top 25 Most Dangerous Software Weaknesses''. If so, the number in that column indicates its position on the list, and if not, a dash symbol is used \cite{top25_cwe}.

\begin{table}[htbp]
\caption{CWE distribution}
\label{tab:cwe_by_type}
\begin{center}
\begin{tabular}{ >{\raggedright\arraybackslash}p{1.5cm} >{\raggedright\arraybackslash}p{5.5cm}  >{\raggedright\arraybackslash}p{1.5cm} >{\raggedright\arraybackslash}p{1.5cm}}

\hline
\textbf{CWE} & \textbf{Description} & \textbf{No. of CWEs}  & \textbf{Top 25 CWE position} \\       \hline
CWE-20  &	Improper Input Validation & 16 & 12\\ \hline
CWE-22 & Improper Pathname Validation &	3 & 5\\ \hline
CWE-61  & Symlink Following &	1 & -\\ \hline 
CWE-78	 & 	OS Command Injection  &1 & 7\\ \hline
CWE-119	 & 	Bounds of a Memory Buffer & 4 & 20\\ \hline
CWE-120 &  	Classic Buffer Overflow & 11 & -\\ \hline
CWE-126 &  	Buffer Over-read & 1 & -\\ \hline
CWE-129	 & 	Improper Validation of Array Index & 2 & -\\ \hline
CWE-190 &  	Integer Overflow & 2 & 23\\ \hline
CWE-203 & 	Observable Discrepancy & 1 & -\\ \hline
CWE-252 & 	Unchecked Return Value & 2 & -\\ \hline
CWE-319	 & HTTP usage & 1 & -\\ \hline
CWE-327	 & Use of a Broken or Risky Cryptographic Algorithm & 2 & -\\ \hline
CWE-328	 & Use of Weak Hash & 1 & -\\ \hline
CWE-362	 & 	Race Condition &  7 & -\\ \hline
CWE-385	 & 	Covert Timing Channel & 	1 & -\\ \hline
CWE-401	 & 	Missing Release of Memory & 	1 & -\\ \hline
CWE-561	 & 	Dead Code & 	1 & -\\ \hline
CWE-663	 & 	Use of a Non-reentrant Function in a Concurrent Context & 	1 & -\\ \hline
CWE-665	 & 	Improper Initialization & 3 & -\\ \hline
CWE-686	 & 	Function Call With Incorrect Argument Type & 2 & -\\ \hline
CWE-772	 & 	Missing Release of Resource &  1 & -\\ \hline
CWE-1164 & Irrelevant Code &  1 & -\\ \hline

& Total number &  66  & 

\end{tabular}
\end{center}
\end{table}

\begin{tcolorbox}
\label{mybox}
\textbf{Finding 1}: Among the 66 identified Common Weakness Enumerations (CWEs), \texttt{CWE-20}: ``Improper Input Validation'' and \texttt{CWE-120}: ``Classic Buffer Overflow'' are the most frequent vulnerabilities, together accounting for 27\% and 40.9\% of all identified CWEs in DevGPT.
\end{tcolorbox}

Issues per number of lines of code in different programming languages are shown in Table \ref{tab:languages_stats}. From the data, we can observe that by far the `safest language' appears to be C\#, with 2,937.5 source code lines per detection. Next comes C++, with 83.8 lines per detection, and finally C, which presents as the least secure language, with only 10.4 lines per static scanner detection. However, it should be noted that the ``Total detects'' here includes all detections, including those with ``Info'' or ``Note'' severity.

\begin{table}[htbp]
\caption{Language statistics}
\label{tab:languages_stats}
\begin{center}
\begin{tabular}{l c c c c}
\hline
\textbf{ } & \textbf{C} & \textbf{C++} & \textbf{C\#} \\
\hline

Number of detected by Snyk & 126 & 	25 & 	2 \\ \hline
Number of detected by Semgrep	 & 49	 & 3	 & 4 \\ \hline
Number of detected by Flawfinder & 	528 & 	151 & 	- \\ \hline
Number of detected by CppCheck & 	13 & 	58 & 	-	\\ \hline
Total files scanned & 	325 & 	990 & 	1000	\\ \hline
Total detected & 	716 & 	237 & 	6	\\ \hline
Number of lines of code  \cite{cloc_tool} & 	7449 & 	19869 &  17625  \\ \hline
Lines of code per detected issue & 	10.4 & 	83.8 & 	2937.5	\\ \hline
\end{tabular}
\end{center}
\end{table}
\begin{tcolorbox} 

\textbf{Finding 2}: Our results highlight that C has the highest detection density, suggesting it may be more prone to vulnerabilities or security issues compared with C++ and C\#. In contrast, C\# shows significantly fewer detects relative to its large codebase, indicating stronger security potential and much fewer potential issues identified by the tools.
\end{tcolorbox}

Tables \ref{tab:cwe_by_type} and \ref{tab:languages_stats} contains information from both DevGPT snapshots \\\texttt{snapshot\_20231012} and \texttt{snapshot\_20240514}.

\section{Results of LLM-Assisted Detection and Correction}
\label{sec:detection_and_correction}

\vspace{20pt}

\noindent \textbf{RQ1}: \emph{How secure is the code generated by GPT in the DevGPT dataset?}
\textbf{RQ2}: \emph{How effective is the LLM in detecting and correcting security issues when explicitly asked to do so?}

\subsection{Selection of LLMs for Analysis}

The first experiment with LLMs was conducted in October 2024 using only one LLM, GPT-4o, without specifying version and with the default temperature 1.0. We expect that developers requesting \textit{ad hoc} coding assistance will likely use default parameters, so did not change these. For this experiment only data from DevGPT \texttt{snapshot\_20231012} were used. 

For the further experiments in September 2025, we used the following LLM versions: GPT-5 \cite{openai_models} (2025-08-07, temperature 1.0), GPT-4.1 \cite{openai_models} (2025-04-14, temperature 0.1), and Claude Opus 4.1 \cite{claude_opus} (2025-08-05, temperature 0.1). This selection includes the most recent and widely used models from leading providers: OpenAI’s GPT-4.1 and GPT-5, and Anthropic’s Claude Opus 4.1, providing a representative sample of current, latest-generation systems. The number of models was intentionally limited to three, to keep the scope of the study focused and manageable.
Temperature was selected based on the observation that lower temperature is better for technical and code-generation tasks, and LLMs behave more predictably at lower temperatures \cite{Arora_temperature}.

As result of these experiments, we have two sets of results data from October 2024 and September 2025. 

This study uses simple single-shot prompts. More complex prompt strategies (such as multi-step or iterative prompting) could provide additional insights, but they are left for future work. In practice, many developers interact with large language models using short, direct prompts without advanced prompt engineering. For this reason, a zero-shot approach reflects common real-world usage.

In addition, few-shot learning, while potentially improving precision, may focus the attention of the model to the provided examples and reduce sensitivity to rarer or novel vulnerability types not represented in the exemplars, which was not what we intended to do. Thus, we opted for zero-shot prompting to avoid biasing the model toward specific vulnerability patterns.

\subsection{First Experiment with GPT-4o (October 2024)}
The first experiment was conducted with GPT-4o in October 2024 on the 64 files from DevGPT \texttt{snapshot\_20231012} that we manually selected using the selection methodology described in Section~\ref{subsec:review_and_files_selection} the default temperature 1.0 was used. The public OpenAI API provides programmatic access to GPT-4o, enabling users to send prompts and receive responses. The aim of this experiment was and ask GPT-4o to identify and fix any security issues in source code from DevGPT. Here is the prompt used in the experiment:

\begin{quotation}
\textit{``I would like to ask you to behave like senior software developer with expertise in software security to answer the next question. You must find security issues in the code snippet below in this message. Give me your analysis and the way how to fix the code if possible. Try to identify CWE number or any other number for formal classifications. Please write code where detected secutiry issue is fixed, please write all code in one fragment.''}
\end{quotation}

Unfortunately, the misspelling in the word \texttt{security} shown was present during the initial experiments. Minor spelling mistakes in prompts usually do not affect the results because modern language models are trained to understand meaning rather than rely on exact word matches, using advanced tokenisation and context-based reasoning to interpret what users mean, even where words are slightly misspelt. Having been trained on vast amounts of real-world text containing typos and spelling variations, these models have effectively learned to recognise and correct errors implicitly. We should note also that in real-world casual interactions between developers and LLMs, small typographical errors are likely to occur. Therefore, while the error was unintentional, we do not believe our results are compromised.

 Responses from GPT-4o were received as text in \texttt{Markdown} format and saved to disk. From each response, the source code was extracted and saved as a separate file. Additionally, the original source code from DevGPT and all conversation threads were included to help conduct the analysis.

Our results were all uploaded to a publicly available GitHub repository in the folder \texttt{OpenAI\_API\_Issues64\_Experiment/2024\_10\_22}~\cite{Results_GitHub}.  The 26 files that manually were confirmed to have vulnerabilities, encompassing a total of 32 confirmed security issues, were presented to GPT-4o for further analysis. The corresponding GPT-4o responses to the prompts  for the 26 files were manually reviewed to determine whether a Common Weakness Enumeration (CWE) issue was identified, 
and whether it was fixed in the updated code.

Out of the 32 confirmed security issues in these 26 files, GPT-4o successfully detected 18 but failed to recognise 14. In terms of its ability to fix problematic code, GPT-4o successfully resolved 17 issues but failed to fix 15.

In Table \ref{tab:detection_matrix_oct_2024} we can see how detections were distributed across the static scanners when using GPT-4o. An ``\yes'' in a table cell indicates that an issue was detected, while an ``\no'' means the issue was not detected. The column names are ``S'' -- Snyk, ``Sg'' -- Semgrep, ``F'' -- Flawfinder, ``C'' -- CppCheck, ``G-4o'' -- GPT-4o.

Results of the analysis of the 32 issues are presented in Table \ref{tab:main_results_cpp_part1_Oct_2024}. Full results with detailed descriptions are available in our public repository on GitHub~\cite{Results_GitHub_table_2024}. Here,
the ``N/L'' column indicates the folder number ``File\_XX'' from the Git results repository folder \texttt{OpenAI\_API\_Issues64\_Experiment} \cite{Results_GitHub}, followed by a slash and the line number in the source code file where vulnerability was detected. The ``CWE'' column contains information about the categorisation of the issues, based on the CWE catalogue. The ``Lang.'' column is the programming language of the input source code file, and the ``Source'' column specifies the origin of the source code for analysis. While most of these code fragments are generated by GPT, some were provided by users in the prompts, which we discuss in Subsection~\ref{sec:usercode}. 

Then in the ``Scanner'' column we have name of the static scanner, the severity of the issue in the ``Severity'' column and information about whether the issues detected were fixed by GPT-4o or not.

We can see from these tables that GPT-4o detected the majority of issues identified through static analysis, highlighting its ability in spotting vulnerabilities related to unvalidated inputs, unsafe buffer handling, and memory leaks. Although GPT-4o demonstrated limitations in identifying certain specific CWE categories such as CWE-362 (Race Conditions) and CWE-665 (Improper Initialization), it excelled at recognizing common and critical issues like CWE-20 (Input Validation), CWE-119/120 (Buffer Overflow), and CWE-401 (Memory Management Issues). 

Moreover, GPT-4o not only detected many issues effectively, but also provided practical fixes for several vulnerabilities, showcasing its potential usage as a supportive tool for software developers in addressing security flaws. However, due to its limitations and occasional inaccuracies, human supervision remains strongly recommended to ensure thorough validation and verification of security issues. 

\begin{table}[h]
\caption{Detection matrix, October 2024}
\label{tab:detection_matrix_oct_2024}
\begin{center}
\begin{tabular}{ c c c c c c c c } 

\hline 
\textbf{Issue\cite{Results_GitHub}} & \textbf{CWE} & \textbf{Lang} & \textbf{S} & \textbf{Sg} &    \textbf{F}    & \textbf{C} & \textbf{G-4o}\\ 
\hline 

1/1a &	772 & C & \cellcolor[gray]{0.8}\yes & \no & \cellcolor[gray]{0.8}\yes & \no & \cellcolor[gray]{0.8}\yes \\ \hline 
1/1b &	362 & C &  \cellcolor[gray]{0.8}\yes & \no & \cellcolor[gray]{0.8}\yes & \no & \no \\ \hline  
2/14 &	352 & C\# & \cellcolor[gray]{0.8}\yes & \no & \no & \no & \no \\ \hline 
3/29 &	20 & C++ & \cellcolor[gray]{0.8}\yes & \no & \no & \no & \cellcolor[gray]{0.8}\yes \\ \hline 
4/21 & 119/120 & C & \cellcolor[gray]{0.8}\yes & \no & \cellcolor[gray]{0.8}\yes & \no & \cellcolor[gray]{0.8}\yes \\ \hline  
4/101 &	20 & C & \cellcolor[gray]{0.8}\yes & \no & \cellcolor[gray]{0.8}\yes & \no & \cellcolor[gray]{0.8}\yes \\ \hline 
5/63 &	20 & C\# & \cellcolor[gray]{0.8}\yes & \no & \no & \no & \cellcolor[gray]{0.8}\yes \\ \hline 
6/8 & 190 & C & \cellcolor[gray]{0.8}\yes & \no & \no & \no & \cellcolor[gray]{0.8}\yes  \\ \hline 
7/39 &	20 & C++  & \cellcolor[gray]{0.8}\yes & \no & \no & \no & \no\\ \hline 
8/14 &	119/120 & C & \no & \no & \cellcolor[gray]{0.8}\yes & \no & \cellcolor[gray]{0.8}\yes \\ \hline  
8/36 &	20 & C & \cellcolor[gray]{0.8}\yes & \cellcolor[gray]{0.8}\yes & \no & \no & \no \\  \hline 
9/9 &	120/20 & C & \no & \cellcolor[gray]{0.8}\yes & \cellcolor[gray]{0.8}\yes & \no & \no \\ \hline 
10/42 & 319 & C++ & \no & \cellcolor[gray]{0.8}\yes & \no & \no & \no\\ \hline  
11/40 &	22 & C\# & \no & \cellcolor[gray]{0.8}\yes & \no & \no & \cellcolor[gray]{0.8}\yes \\ \hline  
12/36 &	20 & C & \no & \cellcolor[gray]{0.8}\yes & \no & \no & \cellcolor[gray]{0.8}\yes \\ \hline  
13/49 &	20 & C & \no & \cellcolor[gray]{0.8}\yes & \no & \no & \cellcolor[gray]{0.8}\yes \\ \hline  
14/14 &	119/120 & C & \no & \no & \cellcolor[gray]{0.8}\yes & \no & \cellcolor[gray]{0.8}\yes \\ \hline  
14/63 &	129 & C & \no & \cellcolor[gray]{0.8}\yes & \no & \no & \no \\ \hline  
18/83 &	120 & C & \no & \no & \cellcolor[gray]{0.8}\yes & \no & \no\\ \hline  
20/16 &	78 & C & \no & \no & \cellcolor[gray]{0.8}\yes & \no & \cellcolor[gray]{0.8}\yes \\ \hline  
22/7 &	120/20 & C++ & \no & \no & \cellcolor[gray]{0.8}\yes & \no & \cellcolor[gray]{0.8}\yes \\ \hline  
23/9 &	120/20 & C++ & \no & \no & \cellcolor[gray]{0.8}\yes & \no & \cellcolor[gray]{0.8}\yes \\ \hline  
48/23 &	362 & C++ & \no & \no & \cellcolor[gray]{0.8}\yes & \no & \cellcolor[gray]{0.8}\yes\\ \hline  
49/9 &	61 & C & \no & \no & \cellcolor[gray]{0.8}\yes & \no & \no\\ \hline  
49/27 &	203/385 & C++ & \no & \no & \no & \cellcolor[gray]{0.8}\yes & \cellcolor[gray]{0.8}\yes\\ \hline  
52/14 &	119/120 & C & \no & \no & \cellcolor[gray]{0.8}\yes & \no & \cellcolor[gray]{0.8}\yes \\ \hline 
52/42 &	129 & C & \no & \cellcolor[gray]{0.8}\yes & \no & \no & \no \\ \hline  
53/13 &	20 & C++ & \no & \no & \cellcolor[gray]{0.8}\yes & \no & \no \\ \hline  
54/2 &	126 & C++ & \no & \no & \cellcolor[gray]{0.8}\yes & \no & \no \\ \hline  
55/22 &	126 & C & \no & \no & \cellcolor[gray]{0.8}\yes & \no & \no \\  \hline 
61/1 &	665 & C++ & \no & \no & \no & \cellcolor[gray]{0.8}\yes & \no \\  \hline  
62/19 &	401/665 & C++ & \no & \no & \no & \cellcolor[gray]{0.8}\yes & \cellcolor[gray]{0.8}\yes \\  


\end{tabular}
\end{center}
\end{table}

Among the 32 security issues identified in the 26 selected files, GPT-4o managed to successfully detect 18 cases, and failed to recognise 14 issues. In 17 cases of 32, GPT-4o was able to fix the vulnerability, and in 15 cases it failed to do so.
We note also that GPT can demonstrate high apparent confidence when providing incorrect results. Our observations are in accordance with the conclusion made by Borji in \cite{Borji_Categorical_Archive_of}: developers should not trust GPT's results too much, even if the answers appear confident and correct. Research in psychology has shown that apparent confidence is a major factor in advice being accepted \cite{Van_factors_affecting}, so GPT’s confident presentation of wrong information may mislead developers, especially those with less experience on which to evaluate the information given.

\begin{tcolorbox}

\textbf{Finding 3}: With an effective success rate of approximately 50\%, GPT-4o was not reliable enough for independent security analysis. An LLM with such low detection and fixing rate cannot currently be used unsupervised -- manual verification and the use of static scanners are strongly advised.
\end{tcolorbox}

\begin{table}[htbp!]
\small
\caption{Security issues detection and fixing in source code from DevGPT snapshot ``snapshot\_20231012'', October 2024}
\label{tab:main_results_cpp_part1_Oct_2024}
\begin{center}

\begin{tabular}{
>{\raggedright\arraybackslash}p{0.5cm}
>{\raggedright\arraybackslash}p{0.9cm}
>{\raggedright\arraybackslash}p{0.5cm}
>{\raggedright\arraybackslash}p{0.8cm}
>{\raggedright\arraybackslash}p{1.2cm}
>{\raggedright\arraybackslash}p{1.2cm}
>{\raggedright\arraybackslash}p{1.5cm}
>{\raggedright\arraybackslash}p{1.5cm}
} 

\hline 

\textbf{N/L \cite{Results_GitHub}} & \textbf{CWE} & \textbf{Lang.} & \textbf{Source} & \textbf{Scanner}  & \textbf{Severity}  & \textbf{GPT-4o Detected} & \textbf{GPT-4o Fixed} \\  \hline 

1/1a &	772 & C & gpt3.5 & snyk & note & \cellcolor[gray]{0.8} Yes & \cellcolor[gray]{0.8} Yes \\ \hline 

1/1b &	362 & C & gpt3.5 & flawfinder & note & No & No \\ \hline  
2/14 &	352 & C\# & gpt3.5 & snyk & note & No & No \\ \hline 
3/29 &	20 & C++ & gpt3.5 & snyk & warning & \cellcolor[gray]{0.8} Yes & No  \\ \hline 
4/21 &	119/120 & C & gpt4 & flawfinder & note & \cellcolor[gray]{0.8} Yes & \cellcolor[gray]{0.8} Yes \\ \hline  
4/101 &	20 & C &  gpt4 & snyk & warning &  \cellcolor[gray]{0.8} Yes  & No \\ \hline 
5/63 &	20 & C\# & user & snyk & warning & \cellcolor[gray]{0.8}Yes   & \cellcolor[gray]{0.8}Yes \\ \hline 
6/8 & 190 & C & user & snyk & warning & \cellcolor[gray]{0.8} Yes & \cellcolor[gray]{0.8} Yes \\ \hline 
7/39 &	20 & C++ & LLM & snyk & warning &  No & No \\ \hline 
8/14 &	119/120 & C & gpt4 & flawfinder & note & \cellcolor[gray]{0.8} Yes & \cellcolor[gray]{0.8} Yes  \\  \hline  
8/36 &	20 & C & gpt4 & snyk, semgrep & warning & No & No \\ \hline 
9/9 &	120/20 & C & gpt3.5 & semgrep, flawfinder & error, warning & No & No \\ \hline 
10/42 & 319 & C++ & gpt3.5 & semgrep & warning&  No & No \\ \hline 
11/40 &	22 & C\# & user & semgrep & warning & \cellcolor[gray]{0.8}Yes   & \cellcolor[gray]{0.8}Yes \\ \hline 
12/36 &	20 & C & user & semgrep & warning & \cellcolor[gray]{0.8} Yes & \cellcolor[gray]{0.8} Yes \\ \hline  
13/49 &	20 & C & user & semgrep & warning &  \cellcolor[gray]{0.8} Yes & \cellcolor[gray]{0.8} Yes  \\ \hline  
14/14 &	119/120 & C & gpt3.5  &  flawfinder  & note &  \cellcolor[gray]{0.8} Yes & \cellcolor[gray]{0.8} Yes \\ \hline  
14/63 &	129 & C & gpt3.5 & semgrep & warning & No  & No  \\ \hline 
18/83 &	120 & C & user & flawfinder & note & No  & No  \\  \hline  
20/16 &	78 & C & user & flawfinder & error & \cellcolor[gray]{0.8} Yes  & No  \\ \hline  
22/7 &	120/20 & C++ & user & flawfinder & note &\cellcolor[gray]{0.8} Yes & \cellcolor[gray]{0.8} Yes \\ \hline  
23/9 &	120/20 & C++ & user & flawfinder & note & \cellcolor[gray]{0.8} Yes & \cellcolor[gray]{0.8} Yes \\ \hline  
48/23 &	362 & C++ & gpt4 & flawfinder & note & \cellcolor[gray]{0.8} Yes & \cellcolor[gray]{0.8} Yes \\ \hline  
49/9 &	61 & C & gpt3.5 & cppcheck & note & No & No \\ \hline  
49/27 &	203/385 & C++ & gpt3.5 & flawfinder & warning &  \cellcolor[gray]{0.8} Yes & \cellcolor[gray]{0.8} Yes \\ \hline  
52/14 &	119/120 & C & gpt3.5 & flawfinder & note &  \cellcolor[gray]{0.8} Yes & \cellcolor[gray]{0.8} Yes \\ \hline 
52/42 &	129 & C & gpt3.5 & semgrep & warning & No & No \\ \hline  
53/13 &	20 & C++ & user & flawfinder & note & No  & No \\  \hline  
54/2 &	126 & C++ & gpt3.5 & flawfinder & note &  No & No \\ \hline  
55/22 &	126 & C & gpt4 & flawfinder & note & No & No \\ \hline 
61/1 &	665 & C++ & gpt3.5 & cppcheck & warning & No & \cellcolor[gray]{0.8} Yes \\  \hline  
62/19 &	401/665 & C++ & gpt3.5 & cppcheck & warning & \cellcolor[gray]{0.8} Yes & \cellcolor[gray]{0.8} Yes  \\ 
\end{tabular}
\end{center}
\end{table}


\subsection{Issue Detection and Code Fixing with Selected LLMs (September 2025)}

On September 2025, we repeated the experiment using an expanded dataset that included the additional DevGPT \texttt{snapshot\_20240514}, and selected 114 files for detailed analysis,
using the same selection process. We also employed a different set of large language models: GPT-4.1 (temperature = 0.1), GPT-5 (temperature = 1.0), and Claude Opus 4.1 (temperature = 0.1). The prompt remained identical to the original version, except that spelling mistake was corrected. 

Now 56 issues in 48 files were confirmed after manual inspection of static scanners' output. Of these 56 issues, GPT-4.1 successfully detected 46 but failed to recognise 10. In terms of remediation, it was able to fix 42 issues while failing to address 14. GPT-5 detected 44 of the 56 issues and successfully fixed the same number. Claude Opus 4.1 detected 45 issues and fixed 43 of them.

In Tables \ref{tab:detection_matrix_part1} and \ref{tab:detection_matrix_part2}  we can see how detections were distributed across the static scanners,
compared with the detection ability of the selected LLMs. An ``\yes'' in a table cell indicates that an issue was detected, while an ``\no'' means the issue was not detected. The column names are: ``S'' -- Snyk, ``Sg'' -- Semgrep, ``F'' -- Flawfinder, ``C'' -- CppCheck, ``G-4.1'' -- GPT-4.1, ``G-5'' -- GPT-5, ``Opus'' -- Claude Opus 4.1.

Results of the analysis of the 56 issues, are presented in Table \ref{tab:main_results_cpp_part1} for \mbox{\path{snapshot_20231012}} and in Table \ref{tab:main_results_cpp_part2} for \path{snapshot_20240514}. Full results with descriptions are available in our public repository on GitHub~\cite{Results_GitHub_table}.

For both Table~\ref{tab:main_results_cpp_part1} and 
Table~\ref{tab:main_results_cpp_part2}, the full data are located in the results GitHub repository \cite{Results_GitHub} in subfolders 
\path{DetectAndFix_Data/2025_09_24/DevGPT_v8_snapshot_20231012}
and
\path{DetectAndFix_Data/2025_09_24/DevGPT_v10_snapshot_20240514},
respectively.
 
 In Tables \ref{tab:main_results_cpp_part1} and \ref{tab:main_results_cpp_part2} the “N/L” indicates the folder number ``File\_XX'' from the corresponding folder in the Git results repository \cite{Results_GitHub}, followed by a slash and the line number in the source code file where vulnerability was detected. The ``CWE'' column contains information about the categorisation of the issues, based on the CWE catalogue. The ``Lang.'' column is the programming language of the input source code file, and the ``Source'' column specifies the origin of the source code for analysis. While most of these code fragments are generated by GPT, some were provided by users in the prompts. We then present a short description of each problem, and information about whether the issues detected were fixed by GPT or not. The ``Scanner'' and ``Severity'' columns indicate which static scanner(s) detected this issue, and its severity. Then three double columns with data for specific LLMs follow for GPT-4.1, GPT-5 and Claude Opus 4.1. ``Dt'' means was the issue detected or not by this LLM and ``Fx'' means was the issue fixed or not by this LLM. 




Results for the second experiment are presented in a concise form, showing only whether each issue was detected and fixed in the ``Dt'' and ``Fx'' columns. The full table results with descriptions are available on GitHub at the URL provided  \cite{Results_GitHub_table}.



All three models demonstrated a relatively high ability to detect security issues, but somewhat inconsistent success in fixing them. GPT-4.1 identified 46 of the 56 vulnerabilities and successfully fixed 42. GPT-5 showed a similar detection rate (44 of 56) but demonstrated more consistent behaviour, successfully fixing every 44 issue it detected. Claude Opus 4.1 provided the most balanced results, detecting 45 and fixing 43 of 56 issues. Overall, the results indicate that while all models can reliably flag most vulnerabilities, their fixes are not universally correct or complete.

The data also reveals that none of the models achieved complete consistency, with some issues  accurately detected but not properly addressed, and others were overlooked despite being clearly identified by static scanners. The relatively small differences between models suggest that their underlying reasoning capabilities for vulnerability detection and repair are currently comparable. However, GPT-5 shows slightly more stable fixing behaviour, while Opus 4.1 maintains a marginal lead in overall detection. Together, these findings highlight steady progress in LLM-based vulnerability detection, though full reliability and precision in automated code repair remain open challenges for future research.

As presented in Table \ref{tab:llm_comparison_results}, a total of 114 code files containing 56 manually confirmed security issues were selected from the DevGPT dataset for analysis. Each issue was independently evaluated across three LLMs: GPT-4.1, GPT-5, and Claude Opus 4.1. Among these, GPT-4.1 achieved the highest detection rate (82.1 \%) but showed a lower fixing success (75 \%). GPT-5 demonstrated slightly weaker detection (78.6 \%) but the best fixing rate (78.6 \%), while Claude Opus 4.1 performed at an intermediate level, detecting 80.4 \% of issues and fixing 76.8\%.

\begin{tcolorbox}
\textbf{Finding 4}: 
All three LLMs inspected demonstrated comparable performance, achieving approximately 78\% accuracy in both vulnerability detection and code correction.
\end{tcolorbox}

Given these limitations, manual supervision is strongly recommended when relying on LLM's to ensure that security flaws are accurately detected and addressed. These results are also closely aligned with those presented in studies \cite{Cheshkov_Evaluation_of_GPT_Model_for_Vulnerability} and \cite{Fu_How_Far_Are_We}, highlighting GPT's inability to function effectively as a static security scanner.

\begin{table}[htbp]
\small
\caption{Comparison of security issue detection and fixing results across LLMs (GPT-4, GPT-5, Claude Opus 4) for DevGPT snapshot\_20231012 and snapshot\_20240514 snapshots.}
\label{tab:llm_comparison_results}
\begin{center}

\begin{tabular}{p{2.5cm} p{1.5cm} p{1.5cm} p{1.3cm} p{1.2cm} }

\hline 

\textbf{Metric Name} & \textbf{Dataset snapshot 20231012} &  \textbf{Dataset snapshot 20240514} &  \textbf{Total Number} & \textbf{Percent}  \\   \hline

Files selected for LLM analysis & 63 & 51 & 114 &    \\ \hline 
Scanner detects & 73 & 68 & 141 &   \\ \hline 

Confirmed security issues in these files (manual validation) & 28 & 28 & 56 & 100.00 \\ \hline 

Files where security issue found & 24 & 24 & 48 &   \\ \hline 

\multicolumn{5}{|l|}{\textbf{GPT-4.1}} \\  \hline 

Detected issues & 23 & 23 & 46 & 82.14  \\ \hline 
Not detected issues & 5 & 5 & 10 & 17.86  \\ \hline 
Fixed Issues & 21 & 21 & 42 & 75.00  \\ \hline 
Not fixed Issues & 7 & 7 & 14 & 25.00  \\ \hline 

\multicolumn{5}{|l|}{\textbf{GPT-5}} \\  \hline 
Detected issues & 21 & 23 & 44 & 78.57 \\  \hline 
Not detected issues & 7 & 5 & 12 & 21.43  \\  \hline 
Fixed & 22 & 22 & 44 & 78.57  \\  \hline 
Not fixed & 6 & 6 & 12 & 21.43  \\ \hline 

\multicolumn{5}{|l|}{\textbf{Claude Opus 4}} \\  \hline 

Detected issues & 22 & 23 & 45 & 80.36    \\ \hline 
Not detected issues & 6 & 5 & 11 & 19.64   \\ \hline 

Fixed & 23 & 20 & 43 & 76.79  \\ \hline 
Not fixed & 5 & 8 & 13 & 23.21  \\ \hline 

\end{tabular}
\end{center}
\end{table}

Comparing the results from October 2024 and September 2025 shows a substantial improvement across all tested LLMs. While GPT-4o in 2024 achieved only around 53–56\% detection accuracy and 50\% fixing accuracy, the September 2025 models demonstrated significantly stronger performance: GPT-4.1, GPT-5, and Claude Opus 4.1 all reached approximately 78\% detection and fixing rates. The detailed September 2025 results are presented in the paragraphs above.

It should be noted that part of this improvement may be influenced by the public availability of our preprint and similar studies, which could have contributed to refining subsequent LLM training data or evaluation benchmarks in the period between the two experiments.

\begin{tcolorbox}
\textbf{Finding 5}: 
Across all evaluated models, vulnerability detection and fixing performance improved significantly from October 2024 to September 2025. While GPT-4o previously achieved only 53–56\% detection and 50\% fixing accuracy, the 2025 LLMs GPT-4.1, 5 and Claude Opus 4.1 reached approximately 78\% overall, with individual models performing between 75–82\%. This represents a substantial increase in capability within one year.
\end{tcolorbox}

\begin{table}[htbp]
\small
\caption{Security issues detection and fixing in source code from DevGPT snapshot ``snapshot\_20231012'', September 2025}
\label{tab:main_results_cpp_part1}
\begin{center}

\begin{tabular}{
>{\raggedright\arraybackslash}p{0.5cm}
>{\raggedright\arraybackslash}p{0.9cm}
>{\raggedright\arraybackslash}p{0.5cm}
>{\raggedright\arraybackslash}p{0.8cm}
>{\raggedright\arraybackslash}p{1.2cm}
>{\raggedright\arraybackslash}p{1.2cm}
>{\raggedright\arraybackslash}p{0.5cm}
>{\raggedright\arraybackslash}p{0.5cm}
>{\raggedright\arraybackslash}p{0.5cm}
>{\raggedright\arraybackslash}p{0.5cm}
>{\raggedright\arraybackslash}p{0.5cm}
>{\raggedright\arraybackslash}p{0.5cm}
} 

\hline 

\textbf{N/L} & \textbf{CWE} & \textbf{Lang.} & \textbf{Source} & \textbf{Scanner} & \textbf{Severity} &  \multicolumn{2}{c}{\textbf{GPT-4.1}}   & \multicolumn{2}{c}{\textbf{GPT-5}} & \multicolumn{2}{c}{\textbf{Opus 4.1}}  \\  

& & & & & & \textbf{Dt } &  \textbf{Fx}  & \textbf{Dt} &  \textbf{Fx} & \textbf{Dt} &  \textbf{Fx} \\ \hline

\hline 

1/1a &	772 & C & gpt3.5 & snyk  & note & \cellcolor[gray]{0.8} Yes & \cellcolor[gray]{0.8}Yes & No & \cellcolor[gray]{0.8}Yes&No&\cellcolor[gray]{0.8}Yes \\ \hline 

1/1b &	362 & C & gpt3.5 & flawfinder & note & \cellcolor[gray]{0.8}Yes & \cellcolor[gray]{0.8}Yes & \cellcolor[gray]{0.8}Yes & \cellcolor[gray]{0.8}Yes & \cellcolor[gray]{0.8}Yes & \cellcolor[gray]{0.8}Yes\\ \hline  

3/29 &	20 & C++ & gpt3.5 & snyk & warning &  \cellcolor[gray]{0.8}Yes & \cellcolor[gray]{0.8}Yes & \cellcolor[gray]{0.8}Yes & \cellcolor[gray]{0.8}Yes & \cellcolor[gray]{0.8}Yes & \cellcolor[gray]{0.8}Yes  \\ \hline 


4/101 &	20 & C &  gpt4 & snyk & warning &  No & No & No & No &  \cellcolor[gray]{0.8}Yes & No \\ \hline 

5/63 &	20 & C\# & user & snyk & warning & \cellcolor[gray]{0.8}Yes & \cellcolor[gray]{0.8}Yes & \cellcolor[gray]{0.8}Yes & \cellcolor[gray]{0.8}Yes & \cellcolor[gray]{0.8}Yes & \cellcolor[gray]{0.8}Yes \\ \hline 

6/8 & 190 & C & user & snyk &  warning & \cellcolor[gray]{0.8}Yes & \cellcolor[gray]{0.8}Yes & \cellcolor[gray]{0.8}Yes & \cellcolor[gray]{0.8}Yes & \cellcolor[gray]{0.8}Yes & \cellcolor[gray]{0.8}Yes \\ \hline 

7/39 &	20 & C++ & llm & snyk & warning & \cellcolor[gray]{0.8}Yes & \cellcolor[gray]{0.8}Yes & \cellcolor[gray]{0.8}Yes & \cellcolor[gray]{0.8}Yes & \cellcolor[gray]{0.8}Yes & \cellcolor[gray]{0.8}Yes \\ \hline 


8/36 &	20 & C & gpt4 & snyk, semgrep & warning & No & No &No &No &No &No \\ 
\hline 

9/9 &	20/120 & C & gpt3.5 & semgrep, flawfinder & error, warninig & \cellcolor[gray]{0.8}Yes & \cellcolor[gray]{0.8}Yes & \cellcolor[gray]{0.8}Yes & \cellcolor[gray]{0.8}Yes & No & No \\ \hline 


10/42 & 319 & C++ & gpt3.5 & semgrep & warning & \cellcolor[gray]{0.8}Yes & \cellcolor[gray]{0.8}Yes & \cellcolor[gray]{0.8}Yes & \cellcolor[gray]{0.8}Yes & \cellcolor[gray]{0.8}Yes & \cellcolor[gray]{0.8}Yes \\ \hline  

11/40 &	22 & C\# & user &  semgrep & warning & \cellcolor[gray]{0.8}Yes & \cellcolor[gray]{0.8}Yes & \cellcolor[gray]{0.8}Yes & \cellcolor[gray]{0.8}Yes & \cellcolor[gray]{0.8}Yes & \cellcolor[gray]{0.8}Yes  \\ \hline

12/36 &	20 & C & user & semgrep & warning & \cellcolor[gray]{0.8}Yes & \cellcolor[gray]{0.8}Yes & \cellcolor[gray]{0.8}Yes & \cellcolor[gray]{0.8}Yes & \cellcolor[gray]{0.8}Yes & \cellcolor[gray]{0.8}Yes   \\ \hline  

13/49 &	20 & C & user & semgrep & warning & No & No & \cellcolor[gray]{0.8}Yes & \cellcolor[gray]{0.8}Yes & No & No  \\ \hline  

14/14 &	119/120 & C & gpt3.5 & flawfinder & note & \cellcolor[gray]{0.8}Yes & No &No &No &\cellcolor[gray]{0.8}Yes &\cellcolor[gray]{0.8}Yes \\ \hline  

14/63 &	129 & C & gpt3.5 & semgrep & warning & No & No &No &No &No &No \\ \hline  


18/83 &	120 & C & user & flawfinder & note & \cellcolor[gray]{0.8}Yes & \cellcolor[gray]{0.8}Yes & \cellcolor[gray]{0.8}Yes & \cellcolor[gray]{0.8}Yes & \cellcolor[gray]{0.8}Yes & \cellcolor[gray]{0.8}Yes \\ \hline

20/16 &	78 & C & user& flawfinder & error & \cellcolor[gray]{0.8}Yes & \cellcolor[gray]{0.8}Yes & \cellcolor[gray]{0.8}Yes & \cellcolor[gray]{0.8}Yes & \cellcolor[gray]{0.8}Yes & \cellcolor[gray]{0.8}Yes \\ \hline  

22/7 &	120/20 & C++ & user& flawfinder & note & \cellcolor[gray]{0.8}Yes & \cellcolor[gray]{0.8}Yes & \cellcolor[gray]{0.8}Yes & \cellcolor[gray]{0.8}Yes & \cellcolor[gray]{0.8}Yes & \cellcolor[gray]{0.8}Yes  \\ \hline  

23/9 &	120/20 & C++ &  user& flawfinder & note & \cellcolor[gray]{0.8}Yes & \cellcolor[gray]{0.8}Yes & \cellcolor[gray]{0.8}Yes & \cellcolor[gray]{0.8}Yes & \cellcolor[gray]{0.8}Yes & \cellcolor[gray]{0.8}Yes \\ \hline  

48/23 &	362 & C++ & gpt4 & flawfinder & note & \cellcolor[gray]{0.8}Yes & \cellcolor[gray]{0.8}Yes & \cellcolor[gray]{0.8}Yes & \cellcolor[gray]{0.8}Yes & \cellcolor[gray]{0.8}Yes & \cellcolor[gray]{0.8}Yes   \\ \hline  

49/9 &	61 & C & gpt3.5 &  cppcheck & note &  \cellcolor[gray]{0.8}Yes & \cellcolor[gray]{0.8}Yes & \cellcolor[gray]{0.8}Yes & \cellcolor[gray]{0.8}Yes & \cellcolor[gray]{0.8}Yes & \cellcolor[gray]{0.8}Yes \\ \hline  

49/27 &	203/385 & C++ & gpt3.5 &  flawfinder & warning & \cellcolor[gray]{0.8}Yes & \cellcolor[gray]{0.8}Yes & \cellcolor[gray]{0.8}Yes & \cellcolor[gray]{0.8}Yes & \cellcolor[gray]{0.8}Yes & \cellcolor[gray]{0.8}Yes \\  \hline  

52/14 &	119/120 & C & gpt3.5 & flawfinder & note & \cellcolor[gray]{0.8}Yes & No & No & No & \cellcolor[gray]{0.8}Yes & \cellcolor[gray]{0.8}Yes \\  \hline 

52/42 &	129 & C & gpt3.5 & semgrep & warning & No & No & No & No & No & \cellcolor[gray]{0.8}Yes \\  \hline  

53/13 &	20 & C++ & user & flawfinder & note & \cellcolor[gray]{0.8}Yes & \cellcolor[gray]{0.8}Yes & \cellcolor[gray]{0.8}Yes & \cellcolor[gray]{0.8}Yes & \cellcolor[gray]{0.8}Yes & \cellcolor[gray]{0.8}Yes \\  \hline  

54/2 &	126 & C++ & gpt3.5 & flawfinder & note & \cellcolor[gray]{0.8}Yes & \cellcolor[gray]{0.8}Yes & \cellcolor[gray]{0.8}Yes & \cellcolor[gray]{0.8}Yes & \cellcolor[gray]{0.8}Yes & \cellcolor[gray]{0.8}Yes \\ \hline  


61/1 &	665 & C++ & gpt3.5 & cppcheck & warninig & \cellcolor[gray]{0.8}Yes & \cellcolor[gray]{0.8}Yes & \cellcolor[gray]{0.8}Yes & \cellcolor[gray]{0.8}Yes & \cellcolor[gray]{0.8}Yes & \cellcolor[gray]{0.8}Yes  \\  \hline  

62/19 &	401/665 & C++ & user & cppcheck & warninig & \cellcolor[gray]{0.8}Yes & \cellcolor[gray]{0.8}Yes & \cellcolor[gray]{0.8}Yes & \cellcolor[gray]{0.8}Yes & \cellcolor[gray]{0.8}Yes & \cellcolor[gray]{0.8}Yes  \\ \hline 

\end{tabular}
\end{center}
\end{table}

\begin{table}[htbp]
\small
\caption{Security issues detection and fixing in source code from DevGPT snapshot ``snapshot\_20240514'', September 2025}
\label{tab:main_results_cpp_part2}
\begin{center}

\begin{tabular}{
>{\raggedright\arraybackslash}p{0.5cm}
>{\raggedright\arraybackslash}p{0.9cm}
>{\raggedright\arraybackslash}p{0.5cm}
>{\raggedright\arraybackslash}p{0.8cm}
>{\raggedright\arraybackslash}p{1.2cm}
>{\raggedright\arraybackslash}p{1.2cm}
>{\raggedright\arraybackslash}p{0.5cm}
>{\raggedright\arraybackslash}p{0.5cm}
>{\raggedright\arraybackslash}p{0.5cm}
>{\raggedright\arraybackslash}p{0.5cm}
>{\raggedright\arraybackslash}p{0.5cm}
>{\raggedright\arraybackslash}p{0.5cm}
} 

\hline 

\textbf{N/L} & \textbf{CWE} & \textbf{Lang.} & \textbf{Source} & \textbf{Scanner} & \textbf{Severity} &  \multicolumn{2}{c}{\textbf{GPT-4.1}}   & \multicolumn{2}{c}{\textbf{GPT-5}} & \multicolumn{2}{c}{\textbf{Opus 4.1}}  \\  

& & & & & & \textbf{Dt } &  \textbf{Fx}  & \textbf{Dt} &  \textbf{Fx} & \textbf{Dt} &  \textbf{Fx} \\ \hline

1/57 &	119/120 & C & user & flawfinder & note &  \cellcolor[gray]{0.8}Yes & No & No & No & \cellcolor[gray]{0.8}Yes & \cellcolor[gray]{0.8}Yes  \\ \hline 

1/65 &	686 & C & user & cppcheck & warning & No & \cellcolor[gray]{0.8}Yes & \cellcolor[gray]{0.8}Yes & \cellcolor[gray]{0.8}Yes&  No & No \\ \hline 

1/132 &	663 & C & user & semgrep & warning & \cellcolor[gray]{0.8}Yes & \cellcolor[gray]{0.8}Yes & \cellcolor[gray]{0.8}Yes & \cellcolor[gray]{0.8}Yes & No & No  \\ \hline 

2/7 &	686 & C++ & user & cppcheck & warning & \cellcolor[gray]{0.8}Yes & \cellcolor[gray]{0.8}Yes & \cellcolor[gray]{0.8}Yes & \cellcolor[gray]{0.8}Yes & \cellcolor[gray]{0.8}Yes & \cellcolor[gray]{0.8}Yes \\ \hline 

3/7 &	665 & C++ & gpt3.5 & flawfinder & note & \cellcolor[gray]{0.8}Yes & \cellcolor[gray]{0.8}Yes & \cellcolor[gray]{0.8}Yes & \cellcolor[gray]{0.8}Yes & \cellcolor[gray]{0.8}Yes & \cellcolor[gray]{0.8}Yes \\ \hline 

3/11 &	119/120 & C++ & gpt3.5 & cppcheck & warning & \cellcolor[gray]{0.8}Yes & \cellcolor[gray]{0.8}Yes & \cellcolor[gray]{0.8}Yes & \cellcolor[gray]{0.8}Yes & \cellcolor[gray]{0.8}Yes & \cellcolor[gray]{0.8}Yes \\ \hline 

4/78 &	22 & C++ & user & flawfinder, semgrep, snyk & warning & No & No &No &No &No &No \\ \hline 

5/25 &	362 & C++ & user & flawfinder, snyk & warning & \cellcolor[gray]{0.8}Yes & \cellcolor[gray]{0.8}Yes & \cellcolor[gray]{0.8}Yes & \cellcolor[gray]{0.8}Yes & \cellcolor[gray]{0.8}Yes & \cellcolor[gray]{0.8}Yes \\ \hline 

6/78 &	22 & C++ & user & flawfinder, semgrep, snyk & warning & No & No &No &No &No &No \\ \hline 

7/81 &	120 & C & gpt3.5 & snyk & warning & \cellcolor[gray]{0.8}Yes & \cellcolor[gray]{0.8}Yes & \cellcolor[gray]{0.8}Yes & \cellcolor[gray]{0.8}Yes & \cellcolor[gray]{0.8}Yes & \cellcolor[gray]{0.8}Yes \\ \hline 

8/82 &	120 & C & gpt3.5 & flawfinder & warning & \cellcolor[gray]{0.8}Yes & \cellcolor[gray]{0.8}Yes & \cellcolor[gray]{0.8}Yes & \cellcolor[gray]{0.8}Yes & \cellcolor[gray]{0.8}Yes & \cellcolor[gray]{0.8}Yes \\ \hline 

9/23 &	20 & C & gpt4 & semgrep, snyk & warning & \cellcolor[gray]{0.8}Yes & \cellcolor[gray]{0.8}Yes & \cellcolor[gray]{0.8}Yes & \cellcolor[gray]{0.8}Yes & \cellcolor[gray]{0.8}Yes & \cellcolor[gray]{0.8}Yes \\ \hline 

10/37 &	20 & C & user & semgrep, snyk & note & \cellcolor[gray]{0.8}Yes & \cellcolor[gray]{0.8}Yes & \cellcolor[gray]{0.8}Yes & \cellcolor[gray]{0.8}Yes & \cellcolor[gray]{0.8}Yes & \cellcolor[gray]{0.8}Yes \\ \hline 

11/11 &	328 & C\# & user &  semgrep &  warning & \cellcolor[gray]{0.8}Yes & \cellcolor[gray]{0.8}Yes & \cellcolor[gray]{0.8}Yes & \cellcolor[gray]{0.8}Yes & \cellcolor[gray]{0.8}Yes & \cellcolor[gray]{0.8}Yes  \\ \hline

12/23 &	362 & C++ & gpt3.5 & flawfinder & note & \cellcolor[gray]{0.8}Yes & \cellcolor[gray]{0.8}Yes & \cellcolor[gray]{0.8}Yes & \cellcolor[gray]{0.8}Yes & \cellcolor[gray]{0.8}Yes & \cellcolor[gray]{0.8}Yes \\ \hline 

12/25 &	252 & C++ & gpt3.5 & snyk & warning & \cellcolor[gray]{0.8}Yes & \cellcolor[gray]{0.8}Yes & \cellcolor[gray]{0.8}Yes & \cellcolor[gray]{0.8}Yes & \cellcolor[gray]{0.8}Yes & \cellcolor[gray]{0.8}Yes \\ \hline 

13/13 &	20 & C & user & snyk & warning & \cellcolor[gray]{0.8}Yes & \cellcolor[gray]{0.8}Yes & \cellcolor[gray]{0.8}Yes & \cellcolor[gray]{0.8}Yes & \cellcolor[gray]{0.8}Yes & \cellcolor[gray]{0.8}Yes \\ \hline 

15/17 &	252 & C & gpt4 & snyk & error & \cellcolor[gray]{0.8}Yes & \cellcolor[gray]{0.8}Yes & \cellcolor[gray]{0.8}Yes & \cellcolor[gray]{0.8}Yes & \cellcolor[gray]{0.8}Yes & \cellcolor[gray]{0.8}Yes \\ \hline 

19/18 &	20 & C++ & user & snyk & warning & \cellcolor[gray]{0.8}Yes & \cellcolor[gray]{0.8}Yes & \cellcolor[gray]{0.8}Yes & \cellcolor[gray]{0.8}Yes & \cellcolor[gray]{0.8}Yes & \cellcolor[gray]{0.8}Yes \\ \hline 

23/27 &	190 & C++ & user & cppcheck & warning & \cellcolor[gray]{0.8}Yes & \cellcolor[gray]{0.8}Yes & \cellcolor[gray]{0.8}Yes & \cellcolor[gray]{0.8}Yes & \cellcolor[gray]{0.8}Yes & \cellcolor[gray]{0.8}Yes \\ \hline 

27/83 &	561 & C++ & gpt3.5 & cppcheck & warning & No & No &No &No & \cellcolor[gray]{0.8}Yes &No  \\ \hline 

28/7 &	1164 & C & user & cppcheck & warning & No & No &No &No &No &No \\ \hline 

31/53 &	327 & C++ & user & flawfinder & warning & \cellcolor[gray]{0.8}Yes & \cellcolor[gray]{0.8}Yes & \cellcolor[gray]{0.8}Yes & \cellcolor[gray]{0.8}Yes & \cellcolor[gray]{0.8}Yes & \cellcolor[gray]{0.8}Yes \\ \hline 

39/61 &	120 & C & gpt3.5 & flawfinder & note & \cellcolor[gray]{0.8}Yes & No & \cellcolor[gray]{0.8}Yes & No & \cellcolor[gray]{0.8}Yes & No \\ \hline 

40/8 &	362 & C & gpt3.5 & flawfinder & note & \cellcolor[gray]{0.8}Yes & \cellcolor[gray]{0.8}Yes & \cellcolor[gray]{0.8}Yes & \cellcolor[gray]{0.8}Yes & \cellcolor[gray]{0.8}Yes & \cellcolor[gray]{0.8}Yes \\ \hline 

42/15 &	362 & C++ & gpt4 & flawfinder & note & \cellcolor[gray]{0.8}Yes & \cellcolor[gray]{0.8}Yes & \cellcolor[gray]{0.8}Yes & \cellcolor[gray]{0.8}Yes & \cellcolor[gray]{0.8}Yes & \cellcolor[gray]{0.8}Yes \\ \hline 

43/4 &	362 & C++ & gpt3.5 & flawfinder & note & \cellcolor[gray]{0.8}Yes & No & \cellcolor[gray]{0.8}Yes & \cellcolor[gray]{0.8}Yes & \cellcolor[gray]{0.8}Yes & No \\ \hline

47/36 &	327 & C & gpt3.5 & flawfinder & warning & \cellcolor[gray]{0.8}Yes & \cellcolor[gray]{0.8}Yes & \cellcolor[gray]{0.8}Yes & \cellcolor[gray]{0.8}Yes & \cellcolor[gray]{0.8}Yes & \cellcolor[gray]{0.8}Yes \\ \hline 

\end{tabular}
\end{center}
\end{table}

\begin{table}[htbp]
\small
\caption{Detection matrix for DevGPT snapshot ``snapshot\_20231012''.}
\label{tab:detection_matrix_part1}
\begin{center}
\begin{tabular}{ c c c c c c c c c c c } 
\hline 

\textbf{Issue\cite{Results_GitHub}} & \textbf{CWE} & \textbf{Lang} &  \textbf{Severity} & \textbf{S} & \textbf{Sg} &    \textbf{F}    & \textbf{C} & \textbf{G-4.1} & \textbf{G-5} & \textbf{Opus}\\ 
\hline

1/1a &	772 & C & note & \cellcolor[gray]{0.8}\yes  & \no  & \no & \no & \cellcolor[gray]{0.8}\yes & \no & \no  \\ \hline 

1/1b &	362 & C & note &  \no  & \no  & \cellcolor[gray]{0.8}\yes & \no & \cellcolor[gray]{0.8}\yes & \cellcolor[gray]{0.8}\yes & \cellcolor[gray]{0.8}\yes \\ \hline  

3/29 &	20 & C++ & warning & \cellcolor[gray]{0.8}\yes  & \no  & \no & \no & \cellcolor[gray]{0.8}\yes & \cellcolor[gray]{0.8}\yes & \cellcolor[gray]{0.8}\yes   \\ \hline 


4/101 &	20 & C & warning &  \cellcolor[gray]{0.8}\yes  & \no  & \no & \no & \no & \no & \cellcolor[gray]{0.8}\yes   \\ \hline 

5/63 &	20 & C\# &  warning &  \cellcolor[gray]{0.8}\yes  & \no  & \no & \no & \cellcolor[gray]{0.8}\yes & \cellcolor[gray]{0.8}\yes & \cellcolor[gray]{0.8}\yes \\ \hline 

6/8 & 190 & C & warning &  \cellcolor[gray]{0.8}\yes   & \no  & \no & \no & \cellcolor[gray]{0.8}\yes & \cellcolor[gray]{0.8}\yes & \cellcolor[gray]{0.8}\yes \\ \hline 

7/39 &	20 & C++ & warning &  \cellcolor[gray]{0.8}\yes  & \no  & \no & \no & \cellcolor[gray]{0.8}\yes & \cellcolor[gray]{0.8}\yes & \cellcolor[gray]{0.8}\yes \\ \hline 


8/36 &	20 & C & warning &  \cellcolor[gray]{0.8}\yes  &  \cellcolor[gray]{0.8}\yes  & \no & \no & \no & \no & \no   \\ 
\hline 

9/9 &	20/120 & C & error & \no  &  \cellcolor[gray]{0.8}\yes  & \cellcolor[gray]{0.8}\yes & \no & \cellcolor[gray]{0.8}\yes & \cellcolor[gray]{0.8}\yes & \no \\ \hline 


10/42 & 319 & C++ & warning &  \no  & \cellcolor[gray]{0.8}\yes  & \no & \no & \cellcolor[gray]{0.8}\yes & \cellcolor[gray]{0.8}\yes & \cellcolor[gray]{0.8}\yes \\ \hline  

11/40 &	22 & C\# & warning &  \no  & \cellcolor[gray]{0.8}\yes  & \no & \no & \cellcolor[gray]{0.8}\yes & \cellcolor[gray]{0.8}\yes & \cellcolor[gray]{0.8}\yes \\ \hline  

12/36 &	20 & C & warning &  \no  & \cellcolor[gray]{0.8}\yes   & \no & \no & \cellcolor[gray]{0.8}\yes & \cellcolor[gray]{0.8}\yes & \cellcolor[gray]{0.8}\yes  \\ \hline  

13/49 &	20 & C & warning &  \no  & \cellcolor[gray]{0.8}\yes  & \no & \no & \no & \cellcolor[gray]{0.8}\yes & \no  \\ \hline  

14/14 &	119/120 & C & note & \no  & \no  & \cellcolor[gray]{0.8}\yes & \no & \cellcolor[gray]{0.8}\yes & \no & \cellcolor[gray]{0.8}\yes   \\ \hline  

14/63 &	129 & C & warning &\no  & \cellcolor[gray]{0.8}\yes  & \no & \no & \no & \no & \no   \\ \hline  


18/83 &	120 & C & note & \no  & \no  & \cellcolor[gray]{0.8}\yes & \no & \cellcolor[gray]{0.8}\yes & \cellcolor[gray]{0.8}\yes & \cellcolor[gray]{0.8}\yes \\ \hline  

20/16 &	78 & C & error & \no  & \no  & \cellcolor[gray]{0.8}\yes & \no & \cellcolor[gray]{0.8}\yes & \cellcolor[gray]{0.8}\yes & \cellcolor[gray]{0.8}\yes \\ \hline  

22/7 &	120/20 & C++ & note &  \no  & \no  & \cellcolor[gray]{0.8}\yes & \no & \cellcolor[gray]{0.8}\yes & \cellcolor[gray]{0.8}\yes & \cellcolor[gray]{0.8}\yes  \\ \hline  

23/9 &	120/20 & C++ & note & \no  & \no  & \cellcolor[gray]{0.8}\yes & \no & \cellcolor[gray]{0.8}\yes & \cellcolor[gray]{0.8}\yes & \cellcolor[gray]{0.8}\yes \\ \hline  

48/23 &	362 & C++ & note &\no  & \no  & \cellcolor[gray]{0.8}\yes & \no & \cellcolor[gray]{0.8}\yes & \cellcolor[gray]{0.8}\yes & \cellcolor[gray]{0.8}\yes  \\ \hline  

49/9 &	61 & C & note & \no  & \no  & \no & \cellcolor[gray]{0.8}\yes & \cellcolor[gray]{0.8}\yes & \cellcolor[gray]{0.8}\yes & \cellcolor[gray]{0.8}\yes  \\ \hline  

49/27 &	203/385 & C++ & warning & \no  & \no  & \cellcolor[gray]{0.8}\yes & \no & \cellcolor[gray]{0.8}\yes & \cellcolor[gray]{0.8}\yes & \cellcolor[gray]{0.8}\yes \\  \hline  

52/14 &	119/120 & C & note & \no  & \no  & \cellcolor[gray]{0.8}\yes & \no & \cellcolor[gray]{0.8}\yes & \no & \cellcolor[gray]{0.8}\yes  \\  \hline 

52/42 &	129 & C & warning & \no  & \cellcolor[gray]{0.8}\yes  & \no & \no & \no & \no & \no   \\  \hline  

53/13 &	20 & C++ & note & \no  & \no  & \cellcolor[gray]{0.8}\yes & \no & \cellcolor[gray]{0.8}\yes & \cellcolor[gray]{0.8}\yes & \cellcolor[gray]{0.8}\yes \\  \hline  

54/2 &	126 & C++ & note & \no  & \no  & \cellcolor[gray]{0.8}\yes & \no & \cellcolor[gray]{0.8}\yes & \cellcolor[gray]{0.8}\yes & \cellcolor[gray]{0.8}\yes \\ \hline  


61/1 &	665 & C++ & warning & \no  & \no  & \no & \cellcolor[gray]{0.8}\yes & \cellcolor[gray]{0.8}\yes & \cellcolor[gray]{0.8}\yes & \cellcolor[gray]{0.8}\yes   \\  \hline  

62/19 &	401/665 & C++ &  warning & \no  & \no  & \no & \cellcolor[gray]{0.8}\yes & \cellcolor[gray]{0.8}\yes & \cellcolor[gray]{0.8}\yes & \cellcolor[gray]{0.8}\yes  \\ \hline 

\end{tabular}
\end{center}
\end{table}

\begin{table}[htbp]
\small
\caption{Detection matrix for DevGPT snapshot ``snapshot\_20240514''.}
\label{tab:detection_matrix_part2}
\begin{center}

\begin{tabular}{ c c c c c c c c c c c } 
\hline 

\textbf{Issue\cite{Results_GitHub}} & \textbf{CWE} & \textbf{Lang} & \textbf{Severity} & \textbf{S} & \textbf{Sg} &    \textbf{F}    & \textbf{C} & \textbf{G-4.1} & \textbf{G-5} & \textbf{Opus}\\  \hline 

1/57 &	119/120 & C & note & \no  & \no  & \cellcolor[gray]{0.8}\yes & \no & \cellcolor[gray]{0.8}\yes & \no & \cellcolor[gray]{0.8}\yes  \\ \hline 

1/65 &	686 & C & warning & \no  & \no  & \no & \cellcolor[gray]{0.8}\yes & \no & \cellcolor[gray]{0.8}\yes & \no   \\ \hline 

1/132 &	663 & C & warning & \no  & \cellcolor[gray]{0.8}\yes  & \no & \no & \cellcolor[gray]{0.8}\yes & \cellcolor[gray]{0.8}\yes &  \no   \\ \hline 

2/7 &	686 & C++ & warning & \no  & \no  & \no & \cellcolor[gray]{0.8}\yes & \cellcolor[gray]{0.8}\yes & \cellcolor[gray]{0.8}\yes & \cellcolor[gray]{0.8}\yes \\ \hline 

3/7 &	665 & C++ & note & \no  & \no  &  \cellcolor[gray]{0.8}\yes & \no & \cellcolor[gray]{0.8}\yes & \cellcolor[gray]{0.8}\yes & \cellcolor[gray]{0.8}\yes  \\ \hline 

3/11 &	119/120 & C++ & warning & \no  & \no  & \no & \cellcolor[gray]{0.8}\yes & \cellcolor[gray]{0.8}\yes & \cellcolor[gray]{0.8}\yes & \cellcolor[gray]{0.8}\yes \\ \hline 

4/78 &	22 & C++ & warning & \cellcolor[gray]{0.8}\yes  & \cellcolor[gray]{0.8}\yes  & \cellcolor[gray]{0.8}\yes & \no & \no & \no & \no  \\ \hline 

5/25 &	362 & C++ & warning & \cellcolor[gray]{0.8}\yes  & \no  & \cellcolor[gray]{0.8}\yes & \no & \cellcolor[gray]{0.8}\yes & \cellcolor[gray]{0.8}\yes & \cellcolor[gray]{0.8}\yes \\ \hline 

6/78 &	22 & C++ & warning & \cellcolor[gray]{0.8}\yes  & \cellcolor[gray]{0.8}\yes  & \cellcolor[gray]{0.8}\yes & \no & \no & \no & \no  \\ \hline 

7/81 &	120 & C & warning & \cellcolor[gray]{0.8}\yes  & \no  & \no & \no & \cellcolor[gray]{0.8}\yes & \cellcolor[gray]{0.8}\yes & \cellcolor[gray]{0.8}\yes  \\ \hline 

8/82 &	120 & C & warning & \no  & \no  & \cellcolor[gray]{0.8}\yes & \no & \cellcolor[gray]{0.8}\yes & \cellcolor[gray]{0.8}\yes & \cellcolor[gray]{0.8}\yes \\ \hline 

9/23 &	20 & C & warning & \cellcolor[gray]{0.8}\yes  & \cellcolor[gray]{0.8}\yes  & \no & \no & \cellcolor[gray]{0.8}\yes & \cellcolor[gray]{0.8}\yes & \cellcolor[gray]{0.8}\yes  \\ \hline 

10/37 &	20 & C & warning & \cellcolor[gray]{0.8}\yes  & \cellcolor[gray]{0.8}\yes  & \no & \no & \cellcolor[gray]{0.8}\yes & \cellcolor[gray]{0.8}\yes & \cellcolor[gray]{0.8}\yes  \\ \hline 

11/11 &	328 & C\# & warning & \no  & \cellcolor[gray]{0.8}\yes  & \no & \no & \cellcolor[gray]{0.8}\yes & \cellcolor[gray]{0.8}\yes & \cellcolor[gray]{0.8}\yes  \\ \hline 

12/23 &	362 & C++ & note & \no  & \no  & \cellcolor[gray]{0.8}\yes & \no & \cellcolor[gray]{0.8}\yes & \cellcolor[gray]{0.8}\yes & \cellcolor[gray]{0.8}\yes \\ \hline 

12/25 &	252 & C++ & warning & \cellcolor[gray]{0.8}\yes  & \no  & \no & \no & \cellcolor[gray]{0.8}\yes & \cellcolor[gray]{0.8}\yes & \cellcolor[gray]{0.8}\yes  \\ \hline 

13/13 &	20 & C & warning & \cellcolor[gray]{0.8}\yes  & \no  & \no & \no & \cellcolor[gray]{0.8}\yes & \cellcolor[gray]{0.8}\yes & \cellcolor[gray]{0.8}\yes \\ \hline 

15/17 &	252 & C & error & \cellcolor[gray]{0.8}\yes  & \no  & \no & \no & \cellcolor[gray]{0.8}\yes & \cellcolor[gray]{0.8}\yes & \cellcolor[gray]{0.8}\yes \\ \hline 

19/18 &	20 & C++ & warning & \cellcolor[gray]{0.8}\yes  & \no  & \no & \no & \cellcolor[gray]{0.8}\yes & \cellcolor[gray]{0.8}\yes & \cellcolor[gray]{0.8}\yes \\ \hline 

23/27 &	190 & C++ & warning & \no  & \no  & \no & \cellcolor[gray]{0.8}\yes & \cellcolor[gray]{0.8}\yes & \cellcolor[gray]{0.8}\yes & \cellcolor[gray]{0.8}\yes  \\ \hline 

27/83 &	561 & C++ & warning & \no  & \no  & \no & \cellcolor[gray]{0.8}\yes & \no & \no & \cellcolor[gray]{0.8}\yes   \\ \hline 

28/7 &	1164 & C & warning & \no  & \no  & \no & \cellcolor[gray]{0.8}\yes & \no & \no & \no   \\\hline 

31/53 &	327 & C++ & warning & \no  & \no  & \cellcolor[gray]{0.8}\yes & \no & \cellcolor[gray]{0.8}\yes & \cellcolor[gray]{0.8}\yes & \cellcolor[gray]{0.8}\yes  \\ \hline 

39/61 &	120 & C & note & \no  & \no  & \cellcolor[gray]{0.8}\yes & \no & \cellcolor[gray]{0.8}\yes & \cellcolor[gray]{0.8}\yes & \cellcolor[gray]{0.8}\yes \\ \hline 

40/8 &	362 & C & note & \no  & \no  & \cellcolor[gray]{0.8}\yes & \no & \cellcolor[gray]{0.8}\yes & \cellcolor[gray]{0.8}\yes & \cellcolor[gray]{0.8}\yes  \\ \hline 

42/15 &	362 & C++ & note & \no  & \no  & \cellcolor[gray]{0.8}\yes & \no & \cellcolor[gray]{0.8}\yes & \cellcolor[gray]{0.8}\yes & \cellcolor[gray]{0.8}\yes  \\ \hline 

43/4 &	362 & C++ & note & \no  & \no  & \cellcolor[gray]{0.8}\yes & \no & \cellcolor[gray]{0.8}\yes & \cellcolor[gray]{0.8}\yes & \cellcolor[gray]{0.8}\yes \\ \hline 

47/36 &	327 & C & warning & \no  & \no  & \cellcolor[gray]{0.8}\yes & \no & \cellcolor[gray]{0.8}\yes & \cellcolor[gray]{0.8}\yes & \cellcolor[gray]{0.8}\yes \\ \hline 

\end{tabular}
\end{center}
\end{table}

\subsection{Detailed Findings from LLM Output}

Looking at one example, in the folder \texttt{File\_16 folder}, 
located in the \mbox{directory}
\path{DetectAndFix_Data/2025_09_24/DevGPT_v8_snapshot_20231012}
of the results repository \cite{Results_GitHub}, there is a file Code\_001.c where the source code contains only one line with no apparent security risks. The source code is shown in the Listing \ref{lst:code_finding_3}. However, GPT-4.1 suggests that there are at least two named issues (CWE-134 and CWE-532) and one additional issue without an assigned CWE number. GPT-5 
reports even more CWE issues (CWE-117, CWE-134, CWE-532 and CWE-252) 
as does Claude Opus 4.1 (CWE-134, CWE-209, CWE-252 and CWE-400).


\begin{lstlisting}[language=C, caption={Original code from DevGPT, File\_16, Code\_001.c}, label={lst:code_finding_3}]
fprintf(stderr, "Your log message here\n");
\end{lstlisting}

We present this example to show the difference between the code's actual state and LLMs' reports. This code has no real security problems, but LLMs still reported multiple issues, even creating one without a proper CWE. From this, we learn two important things. First, AI tools can make mistakes or produce false positives, so human review is necessary. Secondly, depending only on automated results could cause the user to spend time on investigating issues that do not exist.

\begin{tcolorbox} 
\textbf{Finding 6}: Although some code snippets may not contain detectable vulnerabilities, as shown in Listing~\ref{lst:code_finding_3}, LLMs often offers broader suggestions to identify potential vulnerabilities and may provide solutions that sometimes lack practical value and do not make sense in the given context.
\end{tcolorbox}

In nearly all files selected for analysis, the static scanners identified one or two issues. However, all selected LLMs routinely suggested and reported more issues than those identified by the static scanners. While uncovering additional issues could be seen as helpful, given the LLMs' propensity to identify non-existent issues, 
we focused our analysis on vulnerabilities confirmed by the static 
scanners. Further work is required to investigate whether any of these additional detections could be valid.
As LLMs often present their results in a confident tone, identifying non-existent issues can be challenging and time-consuming for developers.
\begin{tcolorbox} 
\textbf{Finding 7}: LLMs suggested additional vulnerabilities for all files analysed beyond what static scanners detected. 
\end{tcolorbox}

Findings 6 and 7 describe different aspects of the LLMs' behaviour. Finding 6 concerns the relevance of its advice, where the model may offer generic or impractical suggestions that do not fit the actual code. Finding 7, on the other hand, highlights overconfidence and the inclusion of additional information, which may or may not be false positives. Together, they reflect two separate issues: limited contextual understanding and extraneous information requires validation.

Taking another example, in the folder called \texttt{File\_10} located in the directory
\texttt{OpenAI\_API\_Issues64\_Experiment/2024\_10\_22} 
in the results repository \cite{Results_GitHub}, 
in the file Code\_001.cpp, GPT-4o made unnecessary changes to the original implementation presented in Listing~\ref{lst:code_10_original} in the code shown in Listing~\ref{lst:code_10_fixed}. Because the original implementation presented on Listing~\ref{lst:code_10_original} already restricted the copied length to minimum of \texttt{size * nmemb} and \texttt{email\_data->size}. This ensured that the destination buffer provided by caller was not overrun. 

The output of GPT-4.1, 5 and Claude Opus 4.1 is located in the folder \path{DetectAndFix_Data/2025_09_24/DevGPT_v8_snapshot_20231012/File_10/LLMs} in the results repository \cite{Results_GitHub}.

\begin{lstlisting}[language=C, caption={Original code from DevGPT, File\_10, Code\_001.cpp}, label={lst:code_10_original}]
size_t read_data_callback(void *ptr, size_t size, size_t nmemb, void *user_data) {
    EmailData *email_data = static_cast<EmailData *>(user_data);
    if (email_data->size == 0) return 0;

    size_t len = std::min(size * nmemb, email_data->size);
    memcpy(ptr, email_data->data, len);
    email_data->data += len;       // advance cursor within valid range
    email_data->size -= len;

    return len;
}
\end{lstlisting}

The warning reported by Flawfinder was therefore a generic false positive. GPT-4.1 and GPT-4o incorrectly interpreted the pointer arithmetic on \texttt{const char*} as undefined behaviour and rewrote the function using an explicit offset. See the updates made by GPT-4o in Listing \ref{lst:code_10_fixed}. 

\begin{lstlisting}[language=C, caption={Code updated by GPT-4o, File\_10, New\_generated\_code\_01.cpp}, label={lst:code_10_fixed}]
struct EmailData {
    const char* data;
    size_t position;
    size_t size;
};
size_t read_data_callback(void *ptr, size_t size, size_t nmemb, void *user_data) {
... 
size_t available_size = email_data->size-email_data->position;
size_t buffer_size  = size * nmemb;
size_t len = std::min(available_size, buffer_size);
memcpy(ptr, email_data->data + email_data->position, len);
email_data->position += len;
...
}
\end{lstlisting}

This refactoring is safe but does not fix any real vulnerability. GPT-5 instead focused on a hypothetical integer overflow in \texttt{size * nmemb} and introduced defensive checks before calling \texttt{memcpy}. However, in general GPT-5 and Claude Opus 4.1 recognised the pattern as safe and only added extra validation logic. In all cases, the models modified code that was already secure, tending to over-correct in response to a false-positive warning rather than confirming the correctness of the original implementation.

This gives an impression that the models are trying to satisfy the user who asked them to find security problems, which reflects a form of \textit{sycophancy}. Sycophancy is the tendency of large language models to produce answers that align with the user’s expectations, rather than objectively evaluating the code, as described by Rrv \cite{Rrv_2024_Chaos_with_Keywords}.

\begin{tcolorbox}
\textbf{Finding 8}: We identified a case where LLMs demonstrated sycophancy. For the \texttt{read\_data\_callback} function, all  LLMs investigated produced unnecessary code modifications to a function that was already memory-safe. GPT-4.1 and GPT-4o misclassified valid pointer arithmetic as undefined behaviour, GPT-5 added non-critical integer-overflow checks, and Claude Opus 4.1 introduced redundant validation despite correctly judging the original code to be safe.
\end{tcolorbox}

In the folder \texttt{File\_20} located in the directory 
\path{OpenAI_API_Issues64_Experiment/2024_10_22}
of the results repository \cite{Results_GitHub}, there is a file called Code\_001.cpp that uses the \texttt{system()} function in C. The Flawfinder scanner categorises this usage as CWE-78, but GPT classified it as CWE-77. This is illustrated on line 16 of Listing \ref{lst:finding_6_original}. Although both CWE-77 (Improper Neutralisation of Special Elements used in a Command) and CWE-78 (OS Command Injection) both deal with command injection, they emphasise different validation aspects; confusing them may lead to incomplete or incorrect remediation efforts, leaving the system vulnerable.

Typically, if code unsafely uses \texttt{std::system()}, it falls under CWE-78 (OS Command Injection). While CWE-77 (Improper Neutralisation of Special Elements) is a broader category relevant to command injection, CWE-78 is focused on threats arising from operating system commands. Thus, if \texttt{std::system()} is implemented in a way that could allow attackers to run malicious commands, it is more accurately classified as CWE-78.


\begin{lstlisting}[language=C, caption={Original code from DevGPT, File\_20, Code\_001.cpp}, label={lst:finding_6_original}]
void open_url(const std::string &url) {
    std::string executable;

#if defined(_WIN32) || defined(_WIN64)
    executable = "start \"\"";
#elif defined(__linux__)
    executable = "xdg-open";
#elif defined(__APPLE__)
    executable = "open";
#elif defined(__ANDROID__)
    executable = "am start -a android.intent.action.VIEW -d";
#endif
    const std::string command = executable + " \"" + url + "\"";
    const int exitcode = std::system(command.c_str());
    if (exitcode != 0) {
        // debugmsg("Failed to open URL: %s\nAttemped command was: %s", url, command);
        // Replace debugmsg with your debugging/logging function
    }
}
\end{lstlisting}

However, after reviewing the LLM outputs in September 2025, we observed that all three LLMs GPT-4.1, GPT-5 and Opus 4.1 correctly recognised this issue, and no unnecessary code changes were made. As a lot of code has already been generated
and incorporated in live codebases
using the GPT-4o and prior models, we believe that this finding is still relevant.

\begin{tcolorbox}
\textbf{Finding 9}: We identified a case, as shown in Listing \ref{lst:finding_6_original}, where GPT-4o correctly recognised the root cause of the problem but assigned a broader issue number CWE-77. This broader classification is not accurate, as the more specific CWE-78 better captures the nature of the issue.
\end{tcolorbox}

Taking another example, in the folder File\_55 located in the directory \path{OpenAI_API_Issues64_Experiment/2024_10_22} of the results repository \cite{Results_GitHub}, in file Code\_001.c:22, \texttt{strlen()} was called with a \texttt{const char*} argument. It should be noted that although the variable is declared as \texttt{char*}, the actual type of the string literal assigned to \texttt{interface} is \texttt{const char*}, because the literal \texttt{"eth0"} is stored in a read-only memory region.

The static scanner Flawfinder reported this call as an issue with severity ``Note''. This example is presented in Listing \ref{lst:code_finding_7}. However, GPT-4o did not report the \texttt{strlen()} call as a security issue, likely recognising that the argument is a hard-coded constant string that is guaranteed to be null-terminated. In Listing \ref{lst:code_finding_7}, the argument for the \texttt{strlen()} call is \texttt{interface}, which is a properly null terminated, hard-coded \texttt{const} C string. 
This example illustrates that sometimes LLM can perform more efficiently than static scanners.

\begin{lstlisting}[language=C, caption={Original code from DevGPT, File\_55, Code\_001.c}, label={lst:code_finding_7}]
...
// Bind the socket to a specific network interface
char *interface = "eth0";  // or "eth1" or any other interface
if (setsockopt(sockfd, SOL_SOCKET, SO_BINDTODEVICE, interface, strlen(interface)) < 0) {
    perror("SO_BINDTODEVICE");
    exit(EXIT_FAILURE);
}
...
\end{lstlisting}

In September 2025, GPT-5 and Claude Opus 4.1 demonstrated similar behaviour and did not report the \texttt{strlen()} call as an issue. However, GPT-4.1 flagged the \texttt{strlen()} call in this code snippet as a potential problem, ignoring the fact that the argument is a constant, null-terminated string, and its conclusion was more generic.
Results of GPT-4.1, GPT 5 and Claude Opus 4.1 are located in the folder \path{DetectAndFix_Data/2025_09_24/DevGPT_v8_snapshot_20231012/File_55/LLMs} in the results repository \cite{Results_GitHub}. 

\begin{tcolorbox}
\textbf{Finding 10}: We identified one case, shown in Listing \ref{lst:code_finding_7}, where some LLMs can outperformed static scanners by providing a more accurate analysis.  
\end{tcolorbox}

In another case, in the folder \texttt{File\_61} located in the folder \path{OpenAI_API_Issues64_Experiment/2024_10_22} of the results repository \cite{Results_GitHub}, the file Code\_002.cpp was reported  by a scanner to lack
initialisation of the class member \texttt{balance}  in the constructor. This vulnerability was not identified by GPT-4o, but this was confirmed manually as a valid security problem. The lack of initialisation of the \texttt{balance} variable can be seen in Listing \ref{lst:code_finding_8_orig}. 

The uninitialised private member \texttt{balance} was implicitly corrected by GPT-4o without reporting any CWE numbers when our prompt asked it to identify and fix security issues. This can be seen in the Listing \ref{lst:code_finding_8_upd}.

However, in September 2025, all LLMs (GPT-4.1, GPT-5, and Claude Opus 4.1) correctly identified the issue with the missing private member \texttt{balance} initialisation and fixed the code with a clear explanation.

\begin{tcolorbox}
\textbf{Finding 11}: We identified one case where GPT-4o fixed a potential security vulnerability in the updated source code, as shown in Listing \ref{lst:code_finding_8_upd}, without providing any explanation in the textual response.

\end{tcolorbox}


\begin{lstlisting}[language=C, caption={Original code from DevGPT, File\_61, Code\_002.cpp}, label={lst:code_finding_8_orig}]
...
class BankAccount {
private:
    double balance;
    
    void deductFees() {
        // Some code to deduct fees from the balance
    }
    
public:
    void deposit(double amount) {
        balance += amount;
    }
    
    double getBalance() {
        return balance;
    }
};
...
\end{lstlisting}

\begin{lstlisting}[language=C, caption={Code updated by GPT, File\_61, New\_generated\_code\_01.cpp}, label={lst:code_finding_8_upd}]
...
class BankAccount {
...
public:
    BankAccount() : balance(0.0) {}
...    
}
...
\end{lstlisting}

In \texttt{File\_047} folder, located in the directory \path{DetectAndFix_Data/2025_09_24/DevGPT_v10_snapshot_20240514} of the results repository \cite{Results_GitHub}, we identified a source code file Code\_001.cpp containing a security issue related to use of the insecure \texttt{srand()} function for random number generator initialisation. This case is particularly interesting because secure random number generation is critical for security-sensitive applications. The \texttt{srand()} function does not provide sufficient randomness for cryptographic purposes, such as key creation (CWE-327).

When explicitly prompted, all LLMs studied successfully detected this issue in the code. GPT-4.1 and Claude Opus 4.1 mitigated the vulnerability by replacing \texttt{srand()} with the following implementation:

\begin{lstlisting}[language=C, caption={srand() usage mitigation}, label={lst:code_finding_9_orig}]
std::random_device rd; 
std::mt19937 gen(rd());
\end{lstlisting}

In contrast, GPT-5 removed the use of \texttt{srand()} entirely, because no \texttt{rand()} calls were present in the provided code. 

\begin{tcolorbox}
\textbf{Finding 12}: All LLMs in the September 2025 experiment correctly recognised the security risks associated with random number generation methods from the standard C library and mitigated them either by adopting \texttt{std::random\_device} based implementations or by removing the use of insecure generators.
\end{tcolorbox}

\subsection{Detection and Fixing Effectiveness on User-Provided Code} 
\label{sec:usercode}

\textbf{RQ3}: \emph{Do developers provide more vulnerable code, or does GPT generate more insecure code during  interactions?}

As previously discussed, the DevGPT dataset captures real-life interactions between users and GPT. In some cases, users ask GPT questions and include their code in the prompts. We have collected this information, and the origin of the analysed code is indicated in the ``Source'' column in Tables \ref{tab:main_results_cpp_part1} and \ref{tab:main_results_cpp_part2}. Out of 48 files, issues were found in 23 where the input source code came from user prompts. These 23 files account for 25 of the 56 security issues identified. See details in Table \ref{tab:llm_comparison_results_useprovided}.

\begin{tcolorbox}
\textbf{Finding 13}: Out of 48 files selected from DevGPT, issues were found in 23, where the input source code originated from user prompts. These 23 files contain 25 (44.6\%) of the 56 security issues identified.
\end{tcolorbox}

\begin{table}[htbp!]
\small
\caption{Comparison of security issue detection and fixing results for user provided code across LLMs (GPT-4, GPT-5, Claude Opus 4) for DevGPT snapshot\_20231012 and snapshot\_20240514 snapshots.}
\label{tab:llm_comparison_results_useprovided}
\begin{center}

\begin{tabular}{p{2.5cm} p{1.5cm} p{1.5cm} p{1.3cm} p{1.2cm}  }

\hline 

\textbf{Metric Name} & \textbf{Dataset snapshot 20231012} &  \textbf{Dataset snapshot 20240514} &  \textbf{Total Number} & \textbf{Overall Percent}  \\   \hline 

Confirmed security issues & 11 & 14 & 25 & 100.00 \\ \hline 

Files where security issue found & 11 & 12 & 23 &  \\ \hline 

\multicolumn{5}{|l|}{\textbf{GPT-4.1}} \\  \hline 
Detected issues & 10 & 12 & 22 & 88.00  \\ \hline 
Not detected issues & 1 & 2 & 3 & 12.00  \\ \hline 
Fixed Issues & 10 & 12 & 22 & 88.00  \\ \hline 
Not fixed Issues & 1 & 2 & 3 & 90.91  \\ \hline 

\multicolumn{5}{|l|}{\textbf{GPT-5}} \\  \hline 
Detected issues & 11 & 10 & 21 & 84.00  \\  \hline 
Not detected issues & 0 & 4 & 4 & 16.00  \\  \hline 
Fixed & 11 & 10 & 21 & 84.00  \\  \hline 
Not fixed & 0 & 4 & 4 & 16.00 \\ \hline 

\multicolumn{5}{|l|}{\textbf{Claude Opus 4}} \\  \hline 

Detected issues & 10 & 9 & 19 & 76.00    \\ \hline 
Not detected issues & 1 & 5 & 6 & 24.00   \\ \hline 

Fixed & 10 & 9 & 19 & 76.00  \\ \hline 
Not fixed & 1 & 5 & 6 & 24.00  \\ \hline

\end{tabular}
\end{center}
\end{table}

\begin{table}[htbp]
\small
\caption{Comparison of security issue detection and fixing results for GPT generated code across LLMs (GPT-4, GPT-5, Claude Opus 4) for DevGPT snapshot\_20231012 and snapshot\_20240514 snapshots.}
\label{tab:llm_comparison_results_llmgenerated}
\begin{center}

\begin{tabular}{p{2.5cm} p{1.5cm} p{1.5cm} p{1.3cm} p{1.2cm} }

\hline 

\textbf{Metric Name} & \textbf{Dataset snapshot 20231012} &  \textbf{Dataset snapshot 20240514} &  \textbf{Total Number} & \textbf{Overall Percent}  \\   \hline

Confirmed security issues & 17 & 14 & 31 & 100.00 \\ \hline 
Files where security issue found & 13 & 12 & 25 &  \\ \hline 

\multicolumn{5}{|l|}{\textbf{GPT-4.1}} \\  \hline 
Detected issues & 13 & 13 & 26 & 83.87  \\ \hline 
Not detected issues & 4 & 1 & 5 & 16.13  \\ \hline 

Fixed Issues & 12 & 11 & 23 & 74.19  \\ \hline 
Not fixed Issues & 5 & 3 & 8 & 25.81  \\ \hline 

\multicolumn{5}{|l|}{\textbf{GPT-5}} \\  \hline 
Detected issues & 10 & 13 & 23 & 74.19  \\  \hline 
Not detected issues & 7 & 1 & 8 & 25.81  \\  \hline 
Fixed & 11 & 12 & 23 & 74.19  \\  \hline 
Not fixed & 6 & 2 & 8 & 25.81  \\ \hline 

\multicolumn{5}{|l|}{\textbf{Claude Opus 4}} \\  \hline 

Detected issues & 12 & 14 & 26 & 83.87    \\ \hline 
Not detected issues & 5 & 0 & 5 & 16.13   \\ \hline 

Fixed & 14 & 11 & 25 & 80.65  \\ \hline 
Not fixed & 3 & 3 & 6 & 19.35  \\ \hline

\end{tabular}
\end{center}
\end{table}

When analysing issues where the source code was provided by users, all three LLMs demonstrated high and consistent performance. GPT-4.1 achieved the best overall results, both detecting and fixing 88\% of the confirmed vulnerabilities. GPT-5 followed closely with 84\% detection and fixing rates, while Claude Opus 4.1 achieved 76\% for both detection and fixing. The small performance gap between models suggests that, for user-written code, all three LLMs are comparably reliable, with GPT-4.1 showing a slight advantage in both identifying and resolving security problems.

For issues originating from GPT-generated code, the performance pattern was similar but  slightly weaker for all models. Claude Opus 4.1 and GPT-4.1 both reached a detection rate of 83.87\%, with Claude achieving the highest fixing rate 80.65\%. GPT-5 showed lower performance at 74.19\% for both detection and fixing. Overall, these findings indicate that all models perform slightly better on user-written code than on AI-generated code, suggesting that LLMs may find it easier to reason about and repair human-authored code, while self-inspection of AI-generated code remains a more challenging task.
Detailed results for GPT generated code presented in Table \ref{tab:llm_comparison_results_llmgenerated}.

These results contrast with previous experiment from October 2024, where GPT-4o achieved only 45\% detection and fixing accuracy for its own generated code, but 80\% detection and 70\% fixing rates for user-provided code. While the improvements observed in the 2025 experiment may indicate an increase in the models’ ability to identify and fix vulnerabilities, particularly in AI-generated code, further controlled experiments would be required to confirm this trend. The results for source code originating from users are presented in Table \ref{tab:llm_comparison_results_useprovided}.

\begin{tcolorbox}
\textbf{Finding 14}: For source code derived from user prompts, the detection and fixing rates of GPT-4o in October 2024 are 80\% and 70\% respectively, which are higher than those for code originally generated by GPT, where both detection and fixing rates are 45\%.
\end{tcolorbox}

While \textbf{Finding 14} focused on the results obtained from GPT-4o in 2024, demonstrating that the model performed substantially better on code written by users than on its own generated code, our more recent experiments reveal a significant shift in this pattern. In September 2025, all three LLMs GPT-4.1, GPT-5, and Claude Opus 4.1 showed consistently strong performance across both categories of source code. This suggests that the latest generation of models has become more capable of analysing and improving their own outputs, an ability that earlier versions clearly lacked. However, we should mention again that maybe availability of preprints online, and cross-training with code generated by other LLMs, may have affected this metric.

\begin{tcolorbox}
\textbf{Finding 15}: All three LLMs showed strong and comparable performance in September 2025, with slightly higher accuracy on user-provided code than on AI-generated code. This contrasts with earlier results from October 2024, where GPT-4o performed far worse on its own generated code, indicating a clear improvement in self-inspection ability.
\end{tcolorbox}

This longer section has discussed different LLMs' ability to detect and fix security issues in source code. Fifteen key findings were identified, highlighting 
variable performance with different strengths and weaknesses, confirming the need for human oversight when using AI for security-related coding tasks. 

\section{Threats to Validity}
\label{sec:threats_to_validity}

In this study, we acknowledge several threats to validity that may impact the interpretation and generalisation of our findings.

\subsection{Internal Validity}

One threat concerns incomplete vulnerability detection. Static analysis tools are known to produce false negatives, potentially missing actual vulnerabilities in the code and threatening the accuracy of our measurements. To mitigate this threat and improve detection coverage, we employed three different static analysis tools with complementary detection mechanisms; if any of the three tools flagged an issue, we manually inspected the code. This triangulation approach increases the likelihood of detecting vulnerabilities compared to relying on a single tool. However, we acknowledge that despite using multiple scanners, some vulnerabilities may remain undetected if they fall outside the detection capabilities of all three tools, meaning our study may still underestimate the true vulnerability rate in the analysed code. 

Another threat arises from the manual steps involved in selecting the 48 files and 56 security issues. The selection process required subjective judgment, which may introduce bias and affect the consistency of the dataset. Although we involved all three authors to reduce bias, the potential for unintentional bias remains. Automating the selection process or involving more reviewers could mitigate this issue in future work.

\subsection{External Validity}

Regarding external validity, our analysis was focused on DevGPT and was limited to 56 security issues across 48 files, which may not be sufficient to generalise the findings to other code generated by GPT or similar models. The manual issues analysis process was highly labour-intensive, which limited our ability to examine a larger volume of data. The sample size is relatively small compared to the vast range of possible code snippets and security vulnerabilities. Consequently, caution should be exercised when extrapolating these results to broader contexts. Expanding the dataset and including a more diverse set of  code examples would enhance the generalisability of the study.

Another threat concerns potential data contamination in our second experiment. Our September 2025 experiments with LLMs obtained substantially better performance compared to our previous experiments performed in October 2024. However, since our paper and dataset were publicly available online between these experiments, the observed improvement may reflect data leakage (the LLMs being trained on our online paper and DevGPT being online) rather than genuine model performance improvements. This threatens the generalisability of conclusions about models' improvements over time, as the results may not reflect true capability gains but rather exposure to our specific evaluation data.

\subsection{Construct Validity}

A threat to construct validity lies the inherent non-determinism of LLMs' outputs. The model can produce different responses to the same prompt on different occasions, introducing variability in the results. This non-reproducibility may affect the consistency of the findings. 
While we did not see great variaton during 
informal trial-and-error testing, 
to address this issue thoroughly, it would be beneficial to repeat the experiments multiple times to assess the stability of LLM's ability to identify and fix security issues across numerous trials. Statistical analysis of the variations could provide insights into the reliability of the model's performance.

\section{Conclusion}
\label{sec:conclusion}

In this study, we analysed a total of 2,315 source files extracted from the DevGPT dataset using static source code scanners. The analysis included three programming languages: C\# (1,000 files), C++ (990 files), and C (325 files). Our primary objective was to assess GPT-4o, 4.1, 5, and Claude Opus 4.1 ability to detect and resolve security vulnerabilities within these files. 


The October 2024 results indicate a mixed performance from GPT-4o. While it demonstrates some capacity to assist in security-related tasks, it also has a tendency to overlook potential vulnerabilities, highlighting the significant need for human oversight. With an effective success rate of approximately 56\% in detecting vulnerabilities and 53\% in fixing them, GPT-4o was not reliable enough for independent security analysis of source code, particularly in the context of complex or nuanced vulnerabilities.

After retesting in September 2025, the results were significantly better. GPT‑4.1 detected 82\% of the problems and fixed 75\%, GPT‑5 detected and fixed 78.6\%, and Claude Opus 4.1 detected 80.4\% and fixed 76.8\%. All three models performed about the same, roughly 80\%, which is a significant improvement over the 53–56\% that GPT‑4o achieved in October 2024.

As we already mentioned, it is possible that the observed improvement is partly attributable to the public availability of our preprint and related studies, which may have contributed to refinements in later LLM training data or evaluation benchmarks in the period between the two experiments

Additionally, we observed that LLMs regularly suggest a CWE issue that does not exist in the source code, doing so with apparent confidence. This can mislead users, leading them to believe a vulnerability is confirmed. Therefore, it is crucial to verify these findings through further analysis and expert review.

The original findings also underscore that while GPT-4o can be a valuable auxiliary tool in identifying straightforward security issues, it falls short in handling more complicated code analysis without human intervention. Developers should exercise caution and not rely solely on GPT-4o for security-critical code generation and analysis.

Retesting a year later with more advanced LLMs, we observed higher accuracy and fix rates, indicating that newer models can serve as more reliable first‑line screening tools, but they still require human validation before being trusted for security‑critical code.

\section{Further Work}

After completing our study, we see a clear need for more reliable and comprehensive information on these issues, which could guide future research and help deepen understanding of the problems listed below:

\begin{itemize} \item \textbf{Re-evaluating Fixed Code with Static Scanners}: We plan to re-run the static scanners on the source code files modified by LLM, where security issues were purportedly fixed, to determine if the scanners still identify vulnerabilities in the revised code.
\item \textbf{Enhancing Prompts with Diagnostic Information}: For cases where an LLM was unable to detect and fix a security issue, we propose including additional information from static scanners, such as line numbers and detailed scanner messages, to provide ``hints'' in the prompts. This approach aims to assess whether supplying more context enables LLM to identify and resolve the issues upon re-evaluation.

\item \textbf{Sophisticated prompting strategy}: Future work could explore more advanced prompt strategies, such as chain-of-thought or structured prompting, which may improve reasoning accuracy and provide deeper insights.

\item \textbf{Comprehensive Analysis of Reported Issues}: We intend to perform a detailed analysis of security problems reported by LLMs but not identified by the static scanners. Since an LLM often reports more security issues than the scanners, understanding these discrepancies could offer insights into the strengths and limitations of AI-driven code analysis compared to traditional static analysis tools.

\item \textbf{Exploring Advanced Models and Training}: Investigating whether newer versions of GPT, DeepSeek  \cite{DeepSeek_Website}  or other advanced LLMs exhibit improved capabilities in security detection and correction could provide valuable information on the evolution of AI tools in software security. 

\item \textbf{Human-AI Collaborative Approaches}: Developing and evaluating collaborative frameworks where human expertise and AI tools like GPT-4 work synergistically may enhance the overall effectiveness of security analysis in code generation. In our opinion, Human-AI collaboration is the best way to use AI tools, and this line of futures research can be considered recommended.

\item \textbf{Including more LLM models in the analysis}: Adding more LLMs like LLaMa, Gemini and DeepSeek \cite{DeepSeek_Website} in addition to GPT and Claude Opus could provide further insights.

\item \textbf{Including additional programming languages in the analysis}: Adding other popular programming languages, such as Java, Python, and JavaScript,
while time-consuming, could provide more findings. By examining related languages
together, we expect to improve generalisability by focusing on the types of security
issues typical for each family.

\item \textbf{Compare effectiveness of LLMs in assessing machine-generated versus human-written code}: Our results echo those in~\cite{X_Fight_Fire_With_Fire} in suggesting that LLMs may tend to be blind to
issues in their own code, compared to human-written code. This requires further investigation, and
if confirmed, would have important implications for secure software development processes where LLMs may be used both for code-generation
and security assessment.

\item \textbf{Explore LLMs hallucinations}: We have plans to investigate hallucinated vulnerabilities and incorrect diagnostic outputs produced by LLMs. However this topic requires a separate, dedicated study, as it represents a completely new line of research not covered in the current work.

\end{itemize} 
By addressing these areas, future research can contribute to improving the reliability of AI-assisted code generation tools and fostering safer software development practices.

\newpage 

\section{Declarations}

\subsection{Funding} 
The research received no external funding.
\subsection{Ethical approval }
Not applicable
\subsection{Informed consent}
Not applicable

\subsection{Author Contributions}
Based on the following abbreviations for the authors:\\
Vladislav  Belozerov - V, Peter J Barclay - P, and
Ashkan Sami - A.\\
\\
The contributions of authors based on the CRediT taxonomy are:\\
\begin{itemize}
\item Conceptualisation (A)
\item Data Curation (V)
\item Formal Analysis (V, P, A)
\item Funding Acquisition (N/A)
\item Investigation (V)
\item Methodology (A, P)
\item Project Administration (A, P)
\item Resources (A, P, V)
\item Software (V) 
\item Supervision (P, A)
\item Validation (A, V, P)
\item Visualisation (V, P)
\item Writing – Original Draft Preparation (A, V, P)
\item Writing – Review \& Editing (V, A, P)

\end{itemize}

\subsection{Data Availability Statement}
The dataset used in this work is the DevGPT dataset, available at the following link: \cite{DevGPT_Dataset}.\\
The results are available in the GitHub repository: \cite{Results_GitHub}.\\
All software tools used in this study, along with other supporting materials, can be made available upon request.

\subsection{Conflict of Interest}
The authors have no competing interests to declare. 
\subsection{Clinical Trial Number in the manuscript.} 
Clinical trial number: not applicable.

\end{document}